\input harvmac
\input graphicx

\def\Title#1#2{\rightline{#1}\ifx\answ\bigans\nopagenumbers\pageno0\vskip1in
\else\pageno1\vskip.8in\fi \centerline{\titlefont #2}\vskip .5in}
%

%
%
\ifx\includegraphics\UnDeFiNeD\message{(NO graphicx.tex, FIGURES WILL BE IGNORED)}
\def\figin#1{\vskip2in}
\else\message{(FIGURES WILL BE INCLUDED)}\def\figin#1{#1}
\fi
\def\Fig#1{Fig.~\the\figno\xdef#1{Fig.~\the\figno}\global\advance\figno
 by1}
%
%
%
%
\def\Ifig#1#2#3#4{
\goodbreak\midinsert
\figin{\centerline{
\includegraphics[width=#4truein]{#3}}}
\narrower\narrower\noindent{\footnotefont
{\bf #1:}  #2\par}
\endinsert
}
%
%
\font\ticp=cmcsc10
\def\subsubsec#1{\noindent{\undertext { #1}}}
\def\undertext#1{$\underline{\smash{\hbox{#1}}}$}

\def\cnl{C^\lam_l}

\def\lam{\lambda}
\def\q{{\bf q}}
\def\x{{\bf x}}

\def\roughly#1{\mathrel{\raise.3ex\hbox{$#1$\kern-.75em\lower1ex\hbox{$\sim$}}}}
\def\calN{{\cal N}}

\def\Ic{{\check I}}
\def\Jc{{\check J}}
%
%
\lref\Eden{R.~J.~Eden, P.~V.~Landshoff, D.~I.~Olive, and J.~C.~Polkinghorne, {\sl The analytic S-matrix,} Cambridge University Press (2002).}
\lref\Lehm{H.L.~Lehmann, ``Analytic properties of scattering amplitudes as functions of momentum transfer," Nuov.\ Cim.\ {\bf 10}, 579 (1958).} 
\lref\KabatTB{
  D.~Kabat and M.~Ortiz,
  ``Eikonal Quantum Gravity And Planckian Scattering,''
  Nucl.\ Phys.\  B {\bf 388}, 570 (1992)
  [arXiv:hep-th/9203082].
}
\lref\tHooultra{
  G.~'t Hooft,
  ``Graviton Dominance in Ultrahigh-Energy Scattering,''
  Phys.\ Lett.\  B {\bf 198}, 61 (1987).
}
\lref\VeVe{
  H.~L.~Verlinde and E.~P.~Verlinde,
  ``Scattering at Planckian energies,''
  Nucl.\ Phys.\  B {\bf 371}, 246 (1992)
  [arXiv:hep-th/9110017].
}
\lref\Juan{
  J.~M.~Maldacena,
  ``The large N limit of superconformal field theories and supergravity,''
  Adv.\ Theor.\ Math.\ Phys.\  {\bf 2}, 231 (1998)
  [Int.\ J.\ Theor.\ Phys.\  {\bf 38}, 1113 (1999)]
  [arXiv:hep-th/9711200].
}
\lref\ACVnew{
  D.~Amati, M.~Ciafaloni and G.~Veneziano,
  ``Towards an S-matrix Description of Gravitational Collapse,''
  JHEP {\bf 0802}, 049 (2008)
  [arXiv:0712.1209 [hep-th]].
}
\lref\FSS{
  S.~B.~Giddings,
  ``Flat-space scattering and bulk locality in the AdS/CFT  correspondence,''
  Phys.\ Rev.\  D {\bf 61}, 106008 (2000)
  [arXiv:hep-th/9907129].
}
\lref\Joe{
  J.~Polchinski,
  ``S-matrices from AdS spacetime,''
  arXiv:hep-th/9901076.
}
\lref\Lenny{
  L.~Susskind,
  ``Holography in the flat space limit,''
  arXiv:hep-th/9901079.
}
\lref\GGP{
  M.~Gary, S.~B.~Giddings and J.~Penedones,
  ``Local bulk S-matrix elements and CFT singularities,''
  arXiv:0903.4437 [hep-th].
}
\lref\GaGiAds{
  M.~Gary and S.~B.~Giddings,
  ``The flat space S-matrix from the AdS/CFT correspondence?,''
  arXiv:0904.3544 [hep-th].
}
\lref\LQGST{
  S.~B.~Giddings,
  ``Locality in quantum gravity and string theory,''
  Phys.\ Rev.\  D {\bf 74}, 106006 (2006)
  [arXiv:hep-th/0604072].
}
\lref\BLOT{N.N.~Bogolubov, A.A.~Logunov, A.I.~Oksak, and I.T.~Todorov, {\sl General principles of quantum field theory}, Kluwer Academic Pub. (Dordrecht, 1990).}
\lref\Heems{
  I.~Heemskerk, J.~Penedones, J.~Polchinski and J.~Sully,
  ``Holography from Conformal Field Theory,''
  arXiv:0907.0151 [hep-th].
}
\lref\GiSr{
  S.~B.~Giddings and M.~Srednicki,
  ``High-energy gravitational scattering and black hole resonances,''
  Phys.\ Rev.\  D {\bf 77}, 085025 (2008)
  [arXiv:0711.5012 [hep-th]].
}
\lref\tHooholo{
  G.~'t Hooft,
  ``Dimensional reduction in quantum gravity,''
  arXiv:gr-qc/9310026.
  }
\lref\Sussholo{
  L.~Susskind,
  ``The World as a hologram,''
  J.\ Math.\ Phys.\  {\bf 36}, 6377 (1995)
  [arXiv:hep-th/9409089].
}
\lref\Jaff{
  A.~M.~Jaffe,
  ``High-Energy Behavior In Quantum Field Theory. I. Strictly Localizable
  Fields,''
  Phys.\ Rev.\  {\bf 158}, 1454 (1967).
}
\lref\Duff{M.~J.~Duff, ``Quantum tree graphs and the Schwarzschild solution," Phys.\ Rev.\  D {\bf 7}, 2317 (1973).}
\lref\Penrose{R. Penrose {\sl unpublished} (1974).}
\lref\GiLi{
  S.~B.~Giddings and M.~Lippert,
  ``Precursors, black holes, and a locality bound,''
  Phys.\ Rev.\  D {\bf 65}, 024006 (2002)
  [arXiv:hep-th/0103231]\semi
  S.~B.~Giddings and M.~Lippert,
  ``The information paradox and the locality bound,''
  Phys.\ Rev.\  D {\bf 69}, 124019 (2004)
  [arXiv:hep-th/0402073].
}
\lref\Martbd{A.~Martin, ``Unitarity and high-energy behavior of scattering amplitudes," Phys. Rev. {\bf 129}, 1432 (1963).}
\lref\Wein{
  S.~Weinberg,
 ``Infrared photons and gravitons,''
  Phys.\ Rev.\  {\bf 140}, B516 (1965).
}
\lref\Page{
  D.~N.~Page,
  ``Information in black hole radiation,''
  Phys.\ Rev.\ Lett.\  {\bf 71}, 3743 (1993)
  [arXiv:hep-th/9306083].
}
\lref\AiSe{
  P.~C.~Aichelburg and R.~U.~Sexl,
  ``On the Gravitational field of a massless particle,''
  Gen.\ Rel.\ Grav.\  {\bf 2}, 303 (1971).
}
\lref\EaGi{
  D.~M.~Eardley and S.~B.~Giddings,
  ``Classical black hole production in high-energy collisions,''
  Phys.\ Rev.\  D {\bf 66}, 044011 (2002)
  [arXiv:gr-qc/0201034].
}
\lref\Wat{G.~N.~Watson, {\sl A treatise on the theory of Bessel functions}, 2nd ed., Cambridge U. Press (Cambridge, 1966).}
\lref\QBHB{
  S.~B.~Giddings,
  ``Quantization in black hole backgrounds,''
  Phys.\ Rev.\  D {\bf 76}, 064027 (2007)
  [arXiv:hep-th/0703116].
}
\lref\AdKo{
  T.~Adachi and T.~Kotani,
  ``An impact parameter representation of the scattering problem,''
  Prog.\ Theor.\ Phys.\  {\bf 39}, 430 (1968); Prog.\ Theor.\ Phys.\  {\bf 39}, 785 (1968).
}
\lref\CoPe{W.~N.~Cottingham and R.~F.~Peierls, ``Impact-parameter expansion of high-energy elastic-scattering amplitudes," Phys.\ Rev.\ {\bf 137}, B147, 1965.}
\lref\AdKounit{ T.~Adachi and T.~Kotani,
  ``Unitarity relation in an impact parameter representation," Prog.\ Theor.\ Phys.\ Suppl.\ {\bf 37, 38}, 297, 1966.}
\lref\Astrorev{
  A.~Strominger,
  ``Les Houches lectures on black holes,''
  arXiv:hep-th/9501071.
}
\lref\WeWi{ V.~Weisskopf and E.~P.~Wigner,
  ``Calculation of the natural brightness of spectral lines on the basis of
  Dirac's theory,''
  Z.\ Phys.\  {\bf 63}, 54 (1930)\semi
  ``On the natural line width in the radiation of the harmonic oscillator,''
  Z.\ Phys.\  {\bf 65}, 18 (1930).
  }
\lref\BEGone{J.~Bros, H.~Epstein, and V.~Glaser, ``Some rigorous analyticity properties of the four-point function in momentum space," Nuov.\ Cim.\ Series X {\bf 31},1265 (1964).}
\lref\BEGtwo{J.~Bros, H.~Epstein, and V.~Glaser, ``A proof of the crossing property for two-particle amplitudes in general quantum field theory," Comm.\ Math.\ Phys.\ {\bf 1}, 240 (1965).}
\lref\GGT{
  M.~Gell-Mann, M.~L.~Goldberger and W.~E.~Thirring,
  ``Use of causality conditions in quantum theory,''
  Phys.\ Rev.\  {\bf 95}, 1612 (1954).
}
\lref\BernKD{
  Z.~Bern, J.~J.~Carrasco, L.~J.~Dixon, H.~Johansson and R.~Roiban,
  ``The Ultraviolet Behavior of N=8 Supergravity at Four Loops,''
  arXiv:0905.2326 [hep-th].
}
\lref\SuTu{M.~Sugawara and A.~Tubis, ``Phase representation of analytic functions," Phys.\ Rev.\   {\bf 130}, 2127 (1963).}
\lref\Kino{T.~Kinoshita, ``Number of subtractions in partial-wave dispersion relations," Phys.\ Rev.\   {\bf 154}, 1438 (1966).}
\lref\GMH{
  S.~B.~Giddings, D.~Marolf and J.~B.~Hartle,
  ``Observables in effective gravity,''
  Phys.\ Rev.\  D {\bf 74}, 064018 (2006)
  [arXiv:hep-th/0512200].
}
\lref\GiMa{
  S.~B.~Giddings and D.~Marolf,
  ``A global picture of quantum de Sitter space,''
  Phys.\ Rev.\  D {\bf 76}, 064023 (2007)
  [arXiv:0705.1178 [hep-th]].
}
\lref\Newt{R.~G.~Newton, {\sl Scattering theory of waves and particles}, McGraw-Hill (New York, 1966).}
\lref\GaGi{
  M.~Gary and S.~B.~Giddings,
  ``Relational observables in 2d quantum gravity,''
  arXiv:hep-th/0612191, Phys.\ Rev.\  D {\bf 75} 104007 (2007).
}
\lref\Sanch{N. Sanchez, ``Scattering of scalar waves from a Schwarzschild black hole," J.\ Math.\ Phys.\ {\bf 17}, 688 (1976).}
\lref\Froiss{
  M.~Froissart,
  ``Asymptotic behavior and subtractions in the Mandelstam representation,''
  Phys.\ Rev.\  {\bf 123}, 1053 (1961).
}
\lref\SGfroiss{
  S.~B.~Giddings,
  ``High energy QCD scattering, the shape of gravity on an IR brane, and  the
  Froissart bound,''
  Phys.\ Rev.\  D {\bf 67}, 126001 (2003)
  [arXiv:hep-th/0203004].
}
\lref\GiTh{
  S.~B.~Giddings and S.~D.~Thomas,
  ``High energy colliders as black hole factories: The end of short  distance
  physics,''
  Phys.\ Rev.\  D {\bf 65}, 056010 (2002)
  [arXiv:hep-ph/0106219].
}
\lref\GiddingsBE{
  S.~B.~Giddings,
 ``(Non)perturbative gravity, nonlocality, and nice slices,''
  Phys.\ Rev.\  D {\bf 74}, 106009 (2006)
  [arXiv:hep-th/0606146].
}
\lref\Venez{
  G.~Veneziano,
  ``String-theoretic unitary S-matrix at the threshold of black-hole
  production,''
  JHEP {\bf 0411}, 001 (2004)
  [arXiv:hep-th/0410166].
}
\lref\HoPo{
  G.~T.~Horowitz and J.~Polchinski,
  ``A correspondence principle for black holes and strings,''
  Phys.\ Rev.\  D {\bf 55}, 6189 (1997)
  [arXiv:hep-th/9612146].
}
\lref\GiddingsSJ{
  S.~B.~Giddings,
  ``Black hole information, unitarity, and nonlocality,''
  Phys.\ Rev.\  D {\bf 74}, 106005 (2006)
  [arXiv:hep-th/0605196].
}
\lref\DiLa{
  S.~Dimopoulos and G.~L.~Landsberg,
  ``Black Holes at the LHC,''
  Phys.\ Rev.\ Lett.\  {\bf 87}, 161602 (2001)
  [arXiv:hep-ph/0106295].
}
\lref\Pascos{
  S.~B.~Giddings,
  ``High-energy black hole production,''
  AIP Conf.\ Proc.\  {\bf 957}, 69 (2007)
  [arXiv:0709.1107 [hep-ph]].
}
\lref\ACVone{
  D.~Amati, M.~Ciafaloni and G.~Veneziano,
  ``Superstring Collisions at Planckian Energies,''
  Phys.\ Lett.\  B {\bf 197}, 81 (1987)\semi 
  D.~Amati, M.~Ciafaloni and G.~Veneziano,
  ``Classical and Quantum Gravity Effects from Planckian Energy Superstring
  Collisions,''
  Int.\ J.\ Mod.\ Phys.\  A {\bf 3}, 1615 (1988).
}
\lref\ACVtwo{
  D.~Amati, M.~Ciafaloni and G.~Veneziano,
  ``Can Space-Time Be Probed Below The String Size?,''
  Phys.\ Lett.\  B {\bf 216}, 41 (1989).
}
\lref\ACVthree{
  D.~Amati, M.~Ciafaloni and G.~Veneziano,
  ``Higher Order Gravitational Deflection And Soft Bremsstrahlung In Planckian
  Energy Superstring Collisions,''
  Nucl.\ Phys.\  B {\bf 347}, 550 (1990).
}
\lref\ACVfour{
  D.~Amati, M.~Ciafaloni and G.~Veneziano,
  ``Effective action and all order gravitational eikonal at Planckian
  energies,''
  Nucl.\ Phys.\  B {\bf 403}, 707 (1993).
}
\lref\GrMe{
 D.~J.~Gross and P.~F.~Mende,
  ``The High-Energy Behavior of String Scattering Amplitudes,''
  Phys.\ Lett.\  B {\bf 197}, 129 (1987)\semi
  D.~J.~Gross and P.~F.~Mende,
  ``String Theory Beyond the Planck Scale,''
  Nucl.\ Phys.\  B {\bf 303}, 407 (1988).
}
\lref\Sold{
  M.~Soldate,
  ``Partial Wave Unitarity and Closed String Amplitudes,''
  Phys.\ Lett.\  B {\bf 186}, 321 (1987).
}
\lref\MuSo{
  I.~J.~Muzinich and M.~Soldate,
  ``High-Energy Unitarity of Gravitation and Strings,''
  Phys.\ Rev.\  D {\bf 37}, 359 (1988).
}
\lref\BaFi{
  T.~Banks and W.~Fischler,
  ``A model for high energy scattering in quantum gravity,''
  arXiv:hep-th/9906038.
}
\lref\CFV{
  M.~Chaichian, J.~Fischer and Yu.~S.~Vernov,
  ``Generalization Of The Froissart-Martin Bounds To Scattering In A Space-Time
  Of General Dimension,''
  Nucl.\ Phys.\  B {\bf 383}, 151 (1992).
}
\lref\GGM{
  S.~B.~Giddings, D.~J.~Gross and A.~Maharana,
  ``Gravitational effects in ultrahigh-energy string scattering,''
  arXiv:0705.1816 [hep-th], to appear in Phys. Rev. D.
}
\lref\MyPe{
  R.~C.~Myers and M.~J.~Perry,
  ``Black Holes In Higher Dimensional Space-Times,''
  Annals Phys.\  {\bf 172}, 304 (1986).
}
\lref\GoRo{
  W.~D.~Goldberger and I.~Z.~Rothstein,
  ``An effective field theory of gravity for extended objects,''
  Phys.\ Rev.\  D {\bf 73}, 104029 (2006)
  [arXiv:hep-th/0409156].
}
\lref\RoGi{
  J.~B.~Gilmore and A.~Ross,
  ``Effective field theory calculation of second post-Newtonian binary
  dynamics,''
  Phys.\ Rev.\  D {\bf 78}, 124021 (2008)
  [arXiv:0810.1328 [gr-qc]].
}
\lref\adams{
  A.~Adams, N.~Arkani-Hamed, S.~Dubovsky, A.~Nicolis and R.~Rattazzi,
 ``Causality, analyticity and an IR obstruction to UV completion,''
  JHEP {\bf 0610}, 014 (2006)
  [arXiv:hep-th/0602178].
}
\lref\BFSS{
  T.~Banks, W.~Fischler, S.~H.~Shenker and L.~Susskind,
  ``M theory as a matrix model: A conjecture,''
  Phys.\ Rev.\  D {\bf 55}, 5112 (1997)
  [arXiv:hep-th/9610043].
}
\lref\DGPP{
  J.~Distler, B.~Grinstein, R.~A.~Porto and I.~Z.~Rothstein,
  ``Falsifying Models of New Physics Via WW Scattering,''
  Phys.\ Rev.\ Lett.\  {\bf 98}, 041601 (2007)
  [arXiv:hep-ph/0604255].
}
\lref\AMV{
N.~Arkani-Hamed, L.~Motl, A.~Nicolis and C.~Vafa,
  JHEP {\bf 0706}, 060 (2007)
  [arXiv:hep-th/0601001].
}
\lref\Martext{A. Martin, ``Extension of the axiomatic analyticity domain of scattering amplitudes by unitarity - I.," Nuov.\ Cim.\ {\bf 42A}, 930 (1966).}
\lref\GiddingsPJ{
  S.~B.~Giddings,
  ``Black holes, information, and locality,''
  Mod.\ Phys.\ Lett.\  A {\bf 22}, 2949 (2007)
  [arXiv:0705.2197 [hep-th]].
}
\lref\SGinfo{S.~B.~Giddings,
  ``Quantum mechanics of black holes,''
  arXiv:hep-th/9412138\semi
  ``The black hole information paradox,''
  arXiv:hep-th/9508151.
}
\lref\GiRy{
  S.~B.~Giddings and V.~S.~Rychkov,
  ``Black holes from colliding wavepackets,''
  Phys.\ Rev.\  D {\bf 70}, 104026 (2004)
  [arXiv:hep-th/0409131].
}
\lref\Boas{R. P. Boas. {\it Entire Functions}. New York, Academic Press, 1954.
}
\lref\kinosh{T. Kinoshita. ``Subtractions in Partial-Wave Dispersion Relations'. Phys. Rev. Lett. 16, 869 (1966)
}
\lref\Martinb{ A. Martin. ``Minimal Interactions at Very High Transfers."  Nuovo Cimento XXXVII N. 2, 671 (1964).
}
\lref\BHMR{
  S.~B.~Giddings,
  ``Black holes and massive remnants,''
  Phys.\ Rev.\  D {\bf 46}, 1347 (1992)
  [arXiv:hep-th/9203059].
}
\Title{
\vbox{
\hbox{CERN-PH-TH/2009-142}}
\vbox{\baselineskip12pt
}}
{\vbox{\centerline{The gravitational S-matrix}
}}
\centerline{{\ticp Steven B. Giddings\footnote{$^\ast$}{Email address: giddings@physics.ucsb.edu} and Rafael A. Porto\footnote{$^\dagger$}{Email address: rporto@physics.ucsb.edu } } }
\centerline{\sl Department of Physics}
\centerline{\sl University of California}
\centerline{\sl Santa Barbara, CA 93106}
\vskip.05in
\centerline {and}
\vskip.05in
\centerline{\sl PH-TH, CERN}
\centerline{\sl 1211 Geneve 23, Switzerland}
\vskip.10in
\centerline{\bf Abstract}
We investigate the hypothesized existence of an S-matrix for gravity, and some of its expected general properties.  We first discuss basic questions regarding existence of such a matrix, including those of infrared divergences and description of asymptotic states.
Distinct scattering behavior occurs in the Born, eikonal, and strong gravity regimes, and we describe aspects of both the partial wave and momentum space amplitudes, and their analytic properties, from these regimes.  Classically the strong gravity region would be dominated by formation of black holes, and we assume its unitary quantum dynamics is described by corresponding resonances.  Masslessness limits some  powerful methods and results that apply to massive theories, though a continuation path implying crossing symmetry plausibly still exists.  Physical properties of gravity suggest nonpolynomial amplitudes,  although crossing and causality constrain (with modest assumptions) this nonpolynomial behavior, particularly requiring a polynomial bound in complex $s$ at fixed physical momentum transfer.  
We explore the hypothesis that such behavior corresponds to a nonlocality intrinsic to gravity, but consistent with unitarity, analyticity, crossing, and causality.

\Date{}

\newsec{Introduction}

In a quantum-gravitational theory where spacetime, locality, {\it etc.~} may not be fundamental concepts, an important question is what  quantities are amenable to quantitative analysis.  In this paper, we will assume that flat space, or something which it closely approximates, is an allowed configuration of the theory.  We will moreover assume that there is an action of its  symmetry group, namely the Poincare group, both on this configuration and on perturbations about it.  This suggests that we can consider states incident from infinity, with given momenta and energies, and study their scattering.  The resulting quantum amplitudes should be summarized in an S-matrix.

One would like to understand what properties are expected of such an S-matrix.  For a quantum theory, unitarity is a given.  Analyticity 
in momenta  and crossing symmetry encode important physical features of S-matrices in quantum field theory (QFT), like causality\refs{\Eden}.
Gravity offers some new features whose role needs to be understood.  Masslessness is first, and causes infrared singularities; these we however envision regulating by working in spacetime dimension $D>4$, or by proper formulation of inclusive amplitudes.  Another is growth of the range of gravity with energy, as is seen for example in growth of the Schwarzschild radius of a black hole formed in a high-energy collision.  An important question is how these new features can be reconciled with the others.  One would also like to understand how these and other physical properties either do or don't manifest themselves in a gravitational S-matrix -- particularly locality and causality.  The latter properties are especially interesting, given that a certain lack of locality  could be part of a mechanism for information to escape a black hole, and thus explain the mysteries surrounding the information paradox.  Yet locality is manifest in low-energy descriptions of nature, and   is a cornerstone of QFT; it is also nontrivially related to causality, which plays an important role in consistency of a theory.

In this paper, we carry out some preliminary investigation of these matters, with particular focus on the ultra-high energy regime.  
We will make the maximal analyticity hypothesis\Eden, where one assumes that the only singularities that appear in the scattering amplitudes are those dictated by unitarity. Our investigations will then focus on the question of what can be learned by combining unitarity, analyticity, crossing and causality together with expected general features of gravity. In spite of the plausibly nonlocal behavior of the gravitational amplitudes that we will explore, we have found no evidence for a lack of harmony between such nonlocality and these basic properties. We thus entertain the possibility that an S-matrix representation of  such nonlocal dynamics exists, which retains the
essential physical features.

The next section will further describe the S-matrix hypothesis, and some issues that must be confronted in its formulation, particularly questions of infrared divergences and asymptotic completeness, and summarizes aspects of exclusive amplitudes and their partial wave expansion.  Section three contains a summary of the different scattering regimes (broadly, Born, eikonal, and strong gravity), and aspects of the physics of each.  Section four focusses on the strong gravity regime, where one expects significant contributions from processes classically described as black hole formation.  We parameterize the corresponding intermediate states as resonances, and investigate their implications for the form of the partial wave amplitudes.  Section five further develops the description of these amplitudes, summarizing our knowledge of the contributions to the phase shifts and their imaginary parts from the different regimes.  Section six overviews some properties of amplitudes in momentum space, some of which can be inferred from 
those  of partial wave amplitudes.  In particular, for both forms of amplitudes, we find strong indications of non-polynomial behavior.  Section seven investigates aspects of analyticity and crossing; the latter is less transparent than in a theory with a mass gap.  Nonetheless, there is an argument for crossing, and this together with causality (plus hermitian analyticity and a smoothness assumption) in turn leads to constraints on non-polynomial growth.  Section eight closes with further discussion of nonpolynomiality, and its connection with the question of locality of the theory.

Study of ultraplanckian collisions in gravity has a long history.  In string theory, this includes  \refs{\ACVone\ACVtwo\ACVthree-\ACVfour} and \GrMe, and other prominent early references are \refs{\Sold,\tHooultra,\MuSo,\BaFi}.  An important question is whether string theory resolves the puzzles of this regime.  In particular, the information paradox suggests a breakdown of locality in this context; while string theory is apparently nonlocal due to string extendedness, it has been argued\refs{\LQGST,\GGM} that this effect does not appear to enter in a central way in the regimes of interest.  In fact, the strong gravitational regime, where classically black holes form, apparently corresponds to a breakdown of the gravitational loop expansion.  Ref.~\refs{\ACVnew} has argued for a possible resummation of string amplitudes that continues into this regime, but we view the apparent need for nonlocal mechanics as well as the absence of clearly relevant stringy effects as suggesting that a new ingredient is instead required for fundamental description of this regime\refs{\LQGST}.  Though a perturbative string description appears insufficient for a complete description, it has been argued that non-perturbative dual formulations such as AdS/CFT\refs{\Juan} will address these problems.  While there has been some progress towards extracting a flat space gravitational S-matrix from 
AdS/CFT\refs{\Joe\Lenny\FSS\GGP-\Heems}, some puzzles remain\refs{\FSS,\GaGiAds} about whether this is possible; one expects similar issues in Matrix theory\refs{\BFSS}. Whether or not it is, we take a more general viewpoint, extending work of \refs{\GiSr}:  whatever theory provides this S-matrix, we would like to characterize its features, and some of those may be rather special  in order to describe gravity.  Moreover, it may be that, as suggested in \refs{\GiddingsSJ}, the need to describe such features is in fact a critical clue to the dynamics of a quantum theory of gravity.

\newsec{The hypothesis of the gravitational S-matrix}

It is natural to expect that the problem of high-energy gravitational scattering in asymptotically flat space can be properly formulated in terms of  the S-matrix.  Here, however, one must grapple with some preliminary issues. 

A first issue is that we don't know a precise description of the quantum numbers of these states.  For example, they could be states of string theory, some other completion of supergravity, or some other theory of gravity.  However, in any case, we expect that the asymptotic states include those corresponding to widely separated individual incident particles, {\it e.g.} electrons, neutrinos, {\it etc.}, in order to match our familiar description of nature.  Or, if the theory were string theory, incident states are string states.  We might have states with other quantum numbers as well.  An example of the latter that is sometimes useful to consider is scattering in Minkowski space that is reached by compactification from higher-dimensions; there, one may have incident particles or strings with conserved Kaluza-Klein charge.  In any of these cases, a nice feature of gravity is that it universally couples to all energy, so we view it as plausible that some important features of gravitational scattering, particularly at high-energy, are independent of this detailed description of the asymptotic states.

A second issue is that, in a perturbative description of gravitons propagating in flat space, gravity suffers from infrared divergences in four dimensions, arising from soft gravitons, and as a consequence one must generalize from the S-matrix to inclusive amplitudes.  While it does not seem inconceivable that this is of fundamental importance, we will assume that it is not.  One reason for this is that QED suffers a similar problem, with the simple resolution through inclusive generalization of the S-matrix, summing over soft photon states.  Moreover, we note that this problem is not present if one works with higher-dimensional gravity.  Specifically, for spacetime dimension $D>4$, soft graviton divergences are not present.  (For $D\geq7$, the total cross-section is finite.)  We have already motivated considering higher-dimensional theories, by including the possibilities of string theory or supergravity, or we may simply think of this as dimensional regularization -- in any case, to avoid this issue we will typically work in $D>4$.

Another issue that plausibly comes closer to being fundamental regards the question of asymptotic completeness of states.  The asymptotic completeness condition\foot{See, {\it e.g.}, chapter 7 of \refs{\BLOT}.} states that the Hilbert space of the theory is equivalent to a Fock space of  asymptotic free particles.  However, there are apparent limitations to such a Fock space description.  An example is the locality bound\refs{\GiLi,\GiddingsSJ,\GiddingsBE} and its N-particle generalizations\refs{\LQGST}.  Specifically, if one considers two particles in wavepackets, which we for example can take to be gaussian with central positions and momenta $x,y$ and $p,q$, these have a field theory description in terms of a Fock space state $\phi_{x,p}\phi_{y,q}|0\rangle$.
However, such a  description must break down when we violate the bound
\eqn\locbd{|x-y|^{D-3}>G|p+q|,}
where $G\sim G_D$, the $D$-dimensional Newton constant.  In this regime, gravity becomes strong, and so limits a Fock space description of the system; this limitation in principle extends to arbitrarily large distances.  One may yet be able to construct an asymptotic description of all states in terms of free-particle states, using further evolution -- if one evolves a state violating \locbd\ backwards in time, it generically ceases violating the bound, and would be expected to resolve itself into well-separated free particles asymptotic from infinity.  Thus, with such a limiting procedure, and a weak form of local Lorentz invariance (in order to describe asymptotic particles with relative boosts), one plausibly describes asymptotics in terms of {\it certain} Fock space states.

In short, we will hypothesize the existence of a gravitational S-matrix, or its inclusive generalization in $D=4$.  While we do not have a complete description of the asymptotic states, we will assume that they include states closely approximating particles that are initially widely-separated, and moreover are allowed to have very large relative momenta.  This starting point amounts to making certain assumptions about a weak notion of locality (asymptotically separated particles) and local Lorentz invariance (large relative boosts allowed for widely separated particles).  However, we will {\it not} necessarily assume that stronger forms of locality and local Lorentz invariance are fundamental in the theory.

For practical purposes, it is often convenient to imagine that the asymptotic states correspond to spinless particles of mass $m$, plus gravitons.  With such a collection of asymptotic states $|\alpha\rangle_{in}$, $|\alpha\rangle_{out}$, (taken to be Heisenberg-picture states)
we expect an S-matrix of the form
\eqn\Sdef{S_{\alpha\beta} = {}_{out}\langle\alpha|\beta\rangle_{in} = \langle\alpha|S|\beta\rangle\ .}
As usual, we separate off the non-trivial part as $S= 1+i{\cal T}$.  

\subsec{Exclusive amplitudes}

Much of this paper's discussion will focus on the simplest non-trivial amplitude of the theory, that for 
exclusive $2\rightarrow2$ scattering.
 Here, the transition matrix element $T$ (in the plane wave limit) is then defined by
\eqn\tmat{\langle p_3,p_4|{\cal T}|p_1,p_2\rangle ={\cal T}_{p_3p_4,p_1p_2} = (2\pi)^D \delta^D(p_1+p_2-p_3-p_4) T(s,t)\ ,}
and is a function of the Mandelstam parameters
\eqn\mandparam{s=-(p_1+p_2)^2 = E^2\ ,\ t=-(p_1-p_3)^2\ ,\ u = -(p_1-p_4)^2\ .}

We expect that important features of the theory are encoded in this amplitude and its analyticity properties. 
Since the graviton is massless, amplitudes are singular at $t=0$, and likewise in other channels; for example, the Born approximation to t-channel exchange gives 
\eqn\born{T_{\rm tree}(s,t)=-8\pi G_D  s^2/t. }
We will consider other aspects of analyticity in section seven.

\subsec{Partial wave expansion}

Unitarity and some other physical features of the amplitude are most clearly formulated by working with the $D$-dimensional partial wave expansion, which is\Sold
\eqn\pwexp{T(s,t) = \psi_\lam s^{2-D/2} \sum_{l=0}^\infty (l+\lam) C_l^\lam(\cos \theta) f_l(s)\ .}
Here $\lam=(D-3)/2$,  
\eqn\psidef{\psi_\lam = 2^{4\lam+3}\pi^\lam \Gamma(\lam)\ ,}
and $C_l^\lam$ are Gegenbauer polynomials, with arguments given by the center-of-mass (CM) scattering angle,
\eqn\scatang{\cos\theta = 1+ {2t\over s-4m^2}\ .}
Note that
\eqn\tutheta{t = (4m^2-s)\sin^2(\theta/2)\ ,\ u=(4m^2-s)\cos^2(\theta/2)\ .}
The inverse relationship to \pwexp\ gives the partial wave coefficients $f_l(s)$ in terms of the matrix element,
\eqn\pwderiv{ f_l(s) = {s^{(D-4)/2}\over \gamma_D \cnl(1)} \int_0^\pi d\theta\, \sin^{D-3}\theta\, \cnl(\cos\theta) T\left[s,(4m^2-s)\sin^2(\theta/2)\right]\ ,}
with 
\eqn\gammaDef{\gamma_D = 2 \Gamma\left({D-2\over 2}\right) (16\pi)^{(D-2)/2}\ .}
The unitarity condition
\eqn\pwunit{{\rm Im} f_l(s)\geq |f_l(s)|^2\ ,}
for real $s\geq0$ can be solved in terms of two real parameters, the phase shift $\delta_l(s)$, and the absorptive coefficients $\beta_l(s)\geq0$:
\eqn\pwsoln{ f_l(s) = {i\over 2} \left[1-e^{2i\delta_l(s) - 2 \beta_l(s)}\right]\ .}

It is important to understand the convergence properties of the partial wave expansion \pwexp.  For a theory with a mass gap, the expansion can be shown to converge in the Lehmann ellipse\Lehm, which 
extends  into the unphysical regime $t>0$, $\cos \theta >1$.  This extension is useful for further constraining amplitudes, {\it e.g.} through the Froissart-Martin \refs{\Froiss,\Martbd} bound.  

Masslessness of gravity alters this behavior.  Let us first ask when the partial wave coefficients \pwderiv\ are well defined.  Specifically, at long-distance/small angle, we have the Born approximation, \born.  This gives a pole at zero angle,
$T\sim 1/\theta^2$, and correspondingly the integral \pwderiv\ only converges for $D>4$.  While other long-distance effects, like soft graviton emission, could modify the amplitude \born, we don't expect them to alter this convergence behavior.

In general, a series of the form \pwexp\ converges in an ellipse with foci at $\cos\theta =\pm1$.  
The existence of the singularity in $T$ at $\theta=0$ indicates that the partial wave expansion  does not converge past $\cos\theta =1$.  Thus, the Lehmann ellipse has collapsed into a line segment along the real axis.  Note that one does expect Im$T(\theta=0)$ to be finite for $D\geq7$.
This follows from the optical theorem (see the Appendix) -- as we have noted, the Born cross section given by \born\ is not infrared divergent for $D\geq7$.  However, this finiteness  does not indicate that the expansion of ${\rm Im}T$ can be continued past this point -- higher derivatives of ${\rm Im}T$ are expected to in general diverge at $\theta=0$.

The failure of convergence of the partial wave expansion in the regime $t>0$ is an impediment to using some of the powerful methods that have been successfully applied in theories with a mass gap.  Nonetheless, we suggest that study of partial wave amplitudes can still be useful for inferring features of scattering.  While we are in particular interested in features of the analytic continuation of $T(s,t)$ to complex values of $s$ and $t$, where convergence of the expansion is problematic, we can exploit the {\it inverse} relation \pwderiv.  Regardless of the convergence of the partial wave expansion, we have argued that \pwderiv\ is convergent for $D\geq 5$.  Thus, if physical considerations imply statements about the behavior of $f_l(s)$, these in turn imply properties of the integrand of \pwderiv, and specifically of $T(s,t)$.

\newsec{Scattering regimes}

In different regions of $s$ and $t$, or $E$ and $l$, we expect differing physical behavior of amplitudes.  A more pictorial way to think of these different regimes is as a function of energy and impact parameter $b$ of the collision -- these are after all often variables controlled experimentally.  While the transformation to impact parameter representation suffers from some complexities, our main focus will be on collisions in the ultrahigh-enegy limit, 
$E \gg M_D$, where $M_D^{D-2} = (2\pi)^{D-4}/(8\pi G_D)$ gives the $D$-dimensional Planck mass.  There, for many purposes, we expect the classical relation
\eqn\ltob{l \sim Eb/2\ ,}
which should approximately hold more generally,
to serve as a useful guide to the physics, though we expect precise statements to be more easily made in terms of the conserved quantities $E$ and $l$.

\Ifig{\Fig\figgraph}{Scattering regimes in an impact-parameter picture; the question marks denote possible model dependence discussed in section 3.3. }{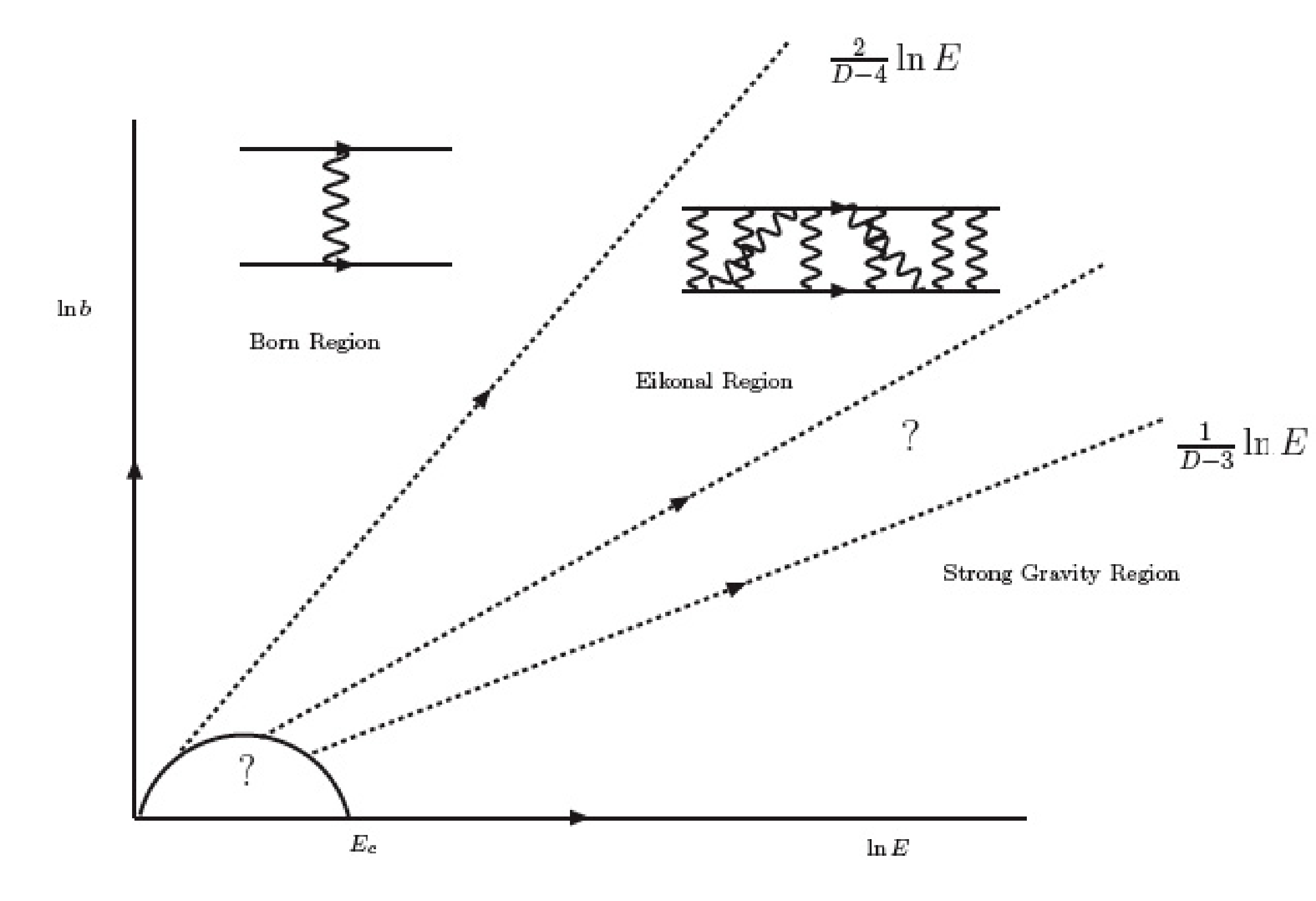}{6}

\figgraph\ illustrates some of the regimes that we expect to be relevant for ultrahigh-energy scattering, in terms of energy and impact parameter.  We will particularly focus on the Born regime, the eikonal regime, and the strong gravity, or ``black hole" regime.

\subsec{Born and eikonal}

The best-understood regime is the Born regime, corresponding to large impact parameters/small angles.  Here, the elastic scattering amplitude, corresponding to single graviton exchange, has been given in \born; one may also consider corrections due to soft graviton emission\refs{\Wein,\ACVthree,\GiSr}.   

As the impact parameter decreases, or the energy increases, diagrams involving exchange of more gravitons become important.  The leading contributions at large impact parameter are the ladder and crossed ladder diagrams, which can be summed to give the eikonal approximation to the amplitude\refs{\ACVone,\ACVtwo,\MuSo,\VeVe,\KabatTB}.\foot{One may inquire about UV divergences of loop diagrams.  However, these are short distance effects, for which we assume there is some UV regulation; for example, string theory might serve this purpose, or even supergravity, if it is perturbatively finite\BernKD.}  This can be written in terms of the {\it eikonal phase}, which arises from a Fourier transformation converting the tree-level amplitude into a function of a variable naturally identified as the impact parameter:
\eqn\eikphase{\eqalign{\chi(x_\perp,s) &= {1\over 2s}\int{d^{D-2}q_\perp\over(2\pi)^{D-2}}\,
e^{-i{\bf q}_\perp\cdot\x_\perp}T_{\rm tree}(s,-q^2_\perp) \cr
&= {4\pi\over (D-4) \Omega_{D-3}} {G_D s\over x_\perp^{D-4}}\ ,}}
where $q_\perp$ is the transverse momentum transfer and 
where 
\eqn\vols{ \Omega_n={2\pi^{(n+1)/2}\over \Gamma[(n+1)/2]}}
is the volume of the unit $n$-sphere.
The eikonal approximation to the amplitude is then
\eqn\Teik{ iT_{\rm eik}(s,t) = 2s \int d^{D-2} x_\perp e^{-i\q_\perp \cdot \x_\perp}(e^{i\chi(x_\perp,s)} -1)\ ,}
expressing the amplitude in an  impact-parameter form.  From \Teik, one sees where eikonal corrections to the Born amplitude become important, namely when the eikonal phase $\chi$ becomes of order one.  Indeed, \refs{\GiSr} showed that at the corresponding point via \ltob, the partial wave phase shifts become of order unity, and thus the eikonal amplitudes unitarize the amplitudes of the Born approximation. (Contributions due to soft graviton emission were also estimated in \GiSr.) In terms of impact parameter, this transition region is given by
\eqn\borntoeik{b \sim {(G_D E^2)}^{1 \over D-4}\ ,}
as is illustrated in \figgraph.  It is alternatively described as the region where the momentum transfer is of order the inverse impact parameter,
\eqn\betrans{\sqrt{-t}\sim {1\over b}\ .}

In general, eikonal approxmations are expected to capture semiclassical physics.  In the high-energy gravitational context, the semiclassical geometry is the collision of two Aichelburg-Sexl shock waves, and various evidence supports the correspondence between \Teik\ and this picture\refs{\tHooultra,\ACVone-\ACVfour}.  In particular, the saddle point of \Teik\ gives a classical scattering angle 
\eqn\classang{\theta_c \sim {1 \over E} {\partial \over \partial b} \chi \sim \left[{R(E) \over b}\right]^{D-3}\ ,}
matching that of a test particle scattering in the Aichelburg-Sexl geometry.  Here, we have introduced the Schwarzschild radius corresponding to the CM energy,
\eqn\schwr{R(E) = {1\over M_D} \left({k_D E\over M_D}\right)^{1/(D-3)},}
where 
\eqn\kddef{k_D = {2(2\pi)^{D-4}\over (D-2) \Omega_{D-2}}.}

\Ifig{\Fig\figH}{The H-diagram, which provides a leading correction to the eikonal amplitudes as scattering angles approach $\theta\sim1$.}{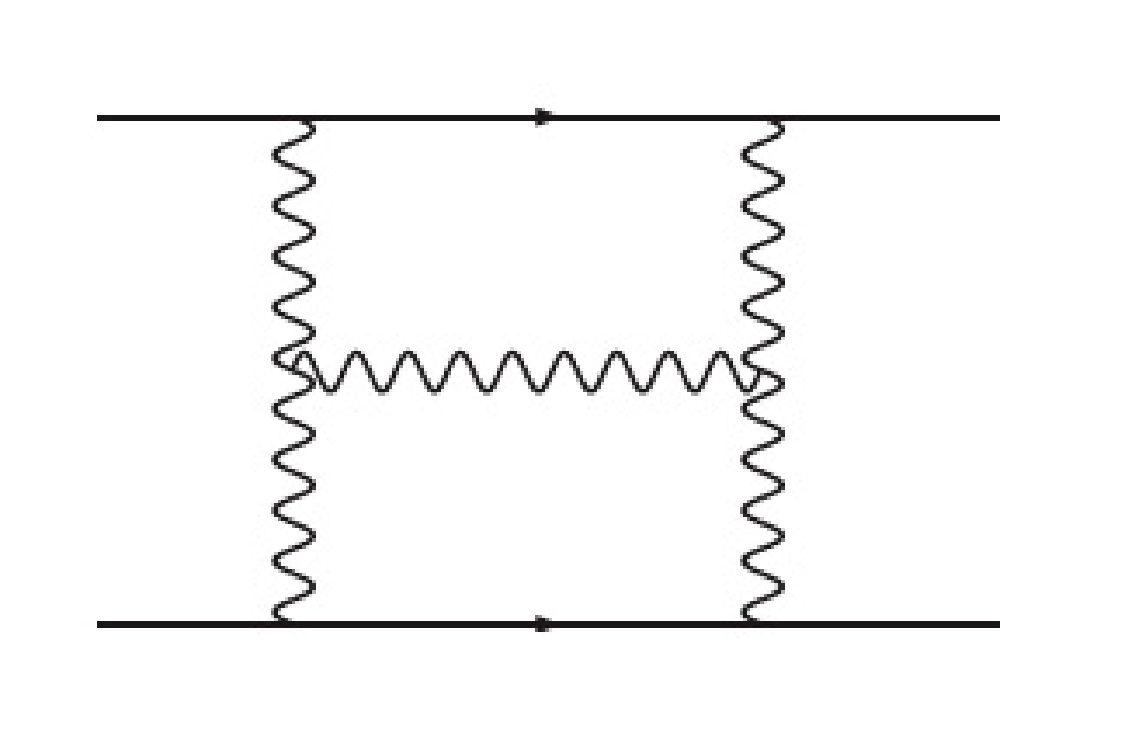}{3}

One finds\ACVfour\ that corrections to the ladder series become important when $\sqrt{-t}\sim E$, or alternatively when the scattering angle reaches $\theta \sim 1$.  Eq.~\classang\ shows that this happens at impact parameter comparable to the Schwarzschild radius, $b\sim R(E)$, as pictured in \figgraph.  A schematic argument for this follows from power-counting.  Consider a diagram arising from a graviton tree  attached to the external lines.  Each graviton vertex  gives a factor $\sqrt {G_D}$.  Those connecting to external lines are accompanied by a $\sqrt s$.  The remaining dimensions come from internal (loop) momenta.  For the processes in question, these have characteristic value\foot{Indeed, in the eikonal regime, the dominant term in the exponential series of \Teik\ occurs at order $N\sim G_D s/b^{D-4}$, corresponding to a characteristic momentum $k\sim \sqrt{-t}/N\sim 1/b$ in each internal line of the $N-1$-loop Feynman ladder diagram.} $k\sim 1/b$.  This counting then produces a power series in $(R/b)^{D-3}$.  A leading such correction, the H-diagram, which has been discussed in \refs{\ACVthree,\ACVfour}, is illustrated in \figH.  One can alternatively understand this expansion by thinking of the external lines as classical sources; using standard power-counting techniques\refs{\GoRo}, one can easily show that the H-diagram is ${\cal O}[(G_DE)^2/r^{2(D-3)}]$ compared to one graviton exchange, if the distance between the sources is $r$ \refs{\RoGi}.  Using $G_DE\sim R^{D-3}$ and taking $r\sim b$ then yields the same expansion parameter. In terms of the semiclassical geometry, at impact parameters $b\sim R$, one forms a trapped surface\refs{\EaGi,\GiRy}, and hence a black hole. 

\subsec{Strong Gravity}

Since corrections to the eikonal amplitudes give terms that differ from the eikonal amplitudes by powers of $[R(E)/b]^{D-3}$,   the region where a classical black hole forms apparently corresponds to a manifest breakdown of the perturbative expansion; it is not even asymptotic.  We can also parameterize this in terms of a critical angular momentum, given by 
\eqn\Ldef{\l\sim L(E) =ER(E)/2\ .}

One might be tempted to believe that a quantum treatment of the evolution can still be given  by performing an expansion in fluctuations about a shifted background -- that of the semiclassical black hole.  However, the problem of the singularity guarantees this is not a complete description.  Moreover, even evolution on spatial ``nice slices" that avoid the singularity is problematical, given that a standard field theory treatment of it leads to the information paradox.\foot{For reviews, see \refs{\SGinfo,\Astrorev}.}  This suggests that the boundary of this regime represents a correspondence boundary, analogous to that for example between classical and quantum mechanics, beyond which local quantum field theory does not give a complete description of the dynamics\QBHB.  In particular, the unitary evolution which we are assuming, in which the quantum information must escape the black hole while it is still comparable to its original radius\refs{\Page}, suggests that the nonperturbative dynamics unitarizing the physics is not local with respect to the semiclassical geometry -- a sort of ``nonlocality principle\refs{\GiddingsSJ,\GiddingsBE}."  (This then fits with the proposed parameterization of part of the correspondence boundary   given by the locality bound \refs{\GiLi,\LQGST,\GiddingsBE}: namely local field theory fails for multi-particle states whose wavefunctions are concentrated inside a radius of size $R(E)$, where $E$ is their combined CM energy.)

While we do not have the means to calculate quantum amplitudes in this regime,\foot{Ref.~\refs{\ACVnew} has suggested analytic continuation of the perturbative sum giving the amplitude in the region $b>R$.  However, one might at best expect such a sum to approximately reconstruct the semiclassical geometry, as in \refs{\Duff}.  Then, in particular, it is not clear how the resulting prescription would give unitary amplitudes that escape the usual reasoning behind the information paradox, which as we have summarized, apparently requires new dynamical ingredients.  Indeed, this paper elaborates on the view that local QFT cannot fully capture the physics of the strong gravitational regime semiclassically associated with black hole formation.}
we can infer some of their properties if we believe that the semiclassical picture of formation of a black hole and its subsequent evaporation provides a good approximate description of the physics when addressing certain coarse-grained questions. Specifically, ref.~\GiSr\ parameterized certain features of the corresponding S-matrix, and we will improve on the corresponding ``black hole ansatz" in subsequent sections.

Of course, investigating the internal dynamics seen, {\it e.g.} by observers falling into black holes, and reconciling that with outside observations such as described by an S-matrix, remains a challenging problem.  Ref.~\QBHB\ has argued for flaws in the ``nice slice argument" for information loss, of two origins.  First, attempts to measure the nice slice state at a level of precision appropriate to investigate information loss lead to large backreaction on the state.  Secondly, fluctuations {\it e.g.} in the Hawking radiation are argued to lead to fluctuations in the nice-slice state after long times.  We expect that sharper investigations should follow from use of proto-local observables\GMH, but ultimately the full non-perturbative dynamics of gravity is plausibly necessary in order to give both a complete picture of infalling observers and of reconciliation of their observations with a unitary S-matrix.

\subsec{Other regimes}

Before turning to further description of the strong gravity regime, it is important to note that at impact parameters larger than $b\sim R(E)$, other features of the dynamics can become relevant.  Indeed, some have argued that this indicates other dynamics besides strong gravity is a dominant feature of high-energy scattering.  To give an example, in the context of string theory, with string mass $M_{st}$, it is possible to make long strings with length $l\sim E/M_{st}^2$.  In fact, such processes are highly suppressed, but \refs{\ACVone} pointed out that such amplitudes receive other important string corrections through ``diffractive excitation" beginning at impact parameters of size $b_t\sim M_D^{-1} (E/M_{st})^{2/(D-2)}$.  Indeed, \refs{\Venez} proposed that this effect may provide important corrections to a picture of black hole formation; if true, this would likely obscure a strong-gravity interpretation of the regime $b\roughly<R(E)$.

Refs.~\refs{\LQGST,\GGM} investigated these effects more closely.  Indeed, as pointed out in \LQGST, a simple picture of the origin of these effects is string excitation arising from the tidal impulse of the gravitational field of the other colliding string.  Moreover, \GGM\ investigated the evolution of the corresponding string states.  For impact parameters $b_t\gg b\gg R(E)$, the asymptotic state of the string is indeed highly excited as a result of this tidal string deformation.  However, for impact parameters $b\roughly<R(E)$, the timescales of horizon formation and string excitation differ significantly.  Roughly, in a semiclassical picture the trapped surface forms before the tidal excitation causes significant extension of the string.  Thus, one seemingly produces a configuration described as a pair of excited strings inside a black hole; in this context there is no clear reason to believe that string extendedness would lead to significant modification of the black hole description of the dynamics.  Likewise, there is not a clear mechanism for string effects to provide the necessary nonlocality with respect to the semiclassical picture, to allow information escape.

Indeed, one can imagine a similar dynamics being relevant for collisions of other composite objects -- hydrogen atoms, protons, etc.  Specifically, when tidal forces reach a size sufficient to excite the internal degrees of freedom of the object, asymptotic states will be excited states.  Thus, there can be model-dependent tidal excitation effects.  However, once impact parameters reach the regime $b\roughly<R(E)$ (and for sufficiently large $E$), such effects are not expected to prevent black hole formation.
Since these model-dependent tidal-excitation effects do not appear to contribute fundamental features to the story, we will largely ignore them in the following discussion.

Another regime that has been of much interest in string theory discussions is that near the string energy, $E\sim M_{st}$, where one might expect to initially see weakly-coupled string excitations.    This region lies in the lower left corner of \figgraph.  One expects such excitations to merge into black holes at a ``correspondence point\refs{\HoPo}" where $R(E_c)\sim 1/M_{st}$.  Our focus will be on higher energies.

\newsec{The strong gravitational regime}

We currently lack a complete quantum description of the strong gravitational regime.  However, we will assume that the quantum description of this regime must be compatible with certain features following from a semiclassical picture of 
 black hole  formation.  If one accepts such a viewpoint, and moreover assumes that the microphysical evolution is unitary, these combined assumptions potentially provide interesting constraints on the dynamics -- particularly in view of the preceding statements that unitary evolution is apparently incompatible with evolution that is local with respect to the semiclassical geometry.

\subsec{Black hole formation}

We begin by recalling basic features of black hole formation in a high-energy collision, which has been extensively studied as a phenomenological feature of models with a low Planck scale\refs{\GiTh,\DiLa}.\foot{For a review with some further references, see \refs{\Pascos}.}

Consider a high-energy collision of two particles, with CM energy $E\gg M_D$.  Let us moreover assume that the wavefunctions of these particles are gaussian wavepackets with characteristic size $\Delta x$, and that these collide with an impact parameter  $b\roughly<R(E)$; for large $E$, we may take $\Delta x\ll R(E)$.  

In the classical description of this process, a trapped surface will form in the geometry\refs{\Penrose,\EaGi}, signaling formation of a black hole, and as a result of the small curvatures, one expects a corresponding statement in a semiclassical approximation to the quantum dynamics\GiRy.  
Not all of the collision energy is trapped in the black hole, which  is initially rather asymmetrical, and radiation (soft gravitons, gauge fields, {\it etc.}) will escape to infinity during the ``balding" process in which it settles down to a Kerr black hole\foot{In models with gauge charges not carried by light particles, the black hole can also carry charge.} of mass $M$. The time scale for balding is of order $\tau_{form}\sim R(E)$, and for impact parameters sufficiently below $R(E)$, the amount of energy lost is an ${\cal O}(1)$ fraction, but not large ({\it e.g.} $\roughly<40\%$), thus $M\approx E$.

Subsequently, the black hole will radiate, initially preferentially radiating states that lower its spin.  The characteristic energy of radiated particles is the Hawking temperature, $T_H\sim 1/R(M)$, and roughly one quantum is emitted per time $\tau \sim R(M)$.

\subsec{Black holes as resonances}

We will thus think of the black holes that form after $\tau_{form}$ as resonances\GiSr.  Since the width for such a state to decay (typically into a lower-energy black hole) is  $\Gamma(M) \sim 1/R(M)$, this is a limit to the sharpness with which we can define the energy of the black hole.  However, black holes with $M\gg M_D$ are sharp resonances in the sense that 
\eqn\sharpcond{ {\Gamma\over M}\sim {1\over RM}\sim {1\over S(M)}\ll 1\ ,}
where $S(M)$ is the Bekenstein-Hawking entropy.

We will assume that the number of possible black hole resonances is given by this entropy.  To be more precise, let us assume that the number of black hole microstates with energies in a range $(M, M+\Delta M)$ is
\eqn\bhnum{\Delta \calN(M) = B(M) e^{S(M)}R(M)\Delta M\ ,}
where $B(M)$ is a possible prefactor that is dimensionless and is expected to have much more slowly-varying energy dependence than the exponential.  Thus the density of black hole states is of the form
\eqn\denstat{
\rho(M) = R B e^{S(M)}\ ,}
and  the total number of states with energy $\leq M$ is $\calN(M)\simeq B(M) \exp\{S(M)\}$.  The spacing between the states is clearly much smaller than their widths.
Let us label the states in the interval $(M, M+1/R)$ as
\eqn\statelab{ |M; I\rangle\ ,}
where $I=1,\cdots,\Delta \calN(M)\sim \exp\{S(M)\}$.  We may further refine the description to project on angular momentum eigenstates, with angular momenta $l$.  In that case, the entropy entering the preceding formulas is expected to be
\eqn\rotS{S(M,l)= {4\pi E R(M,l)\over D-2},}
where $R(E,l)$ is given by\refs{\MyPe}
\eqn\REl{R^{D-5}\left[R^2 + {(D-2)^2l^2\over 4M^2}\right] = {16\pi G_D M\over (D-2) \Omega_{D-2}}.}
For small $l$, this gives an expansion of the form
\eqn\Sexp{S(M,l) = S(M,0)\left( 1- {\rm const.} {l^2\over L^2}\right).}

\subsec{Black hole spectrum and evolution}

Let us explore in more detail the quantum states formed in a collision, which could be either a two-particle collision with a CM energy $\approx E$, or an $n$-body collision.  Note that one can also form a black hole of mass $M$ by producing a higher-mass black hole in a collision with $E\gg M$, and then waiting for that black hole to evaporate to $M$.

Consider general initial multi-particle (but not black hole) states; these can be labeled  by energy, momentum, generalized partial waves, and asymptotic species and spin content.  
Let us work in the CM frame, and ignore the effects of particle spin. Some subset of the  states, denoted $|E;a\rangle_{in}$, will form a black hole; examples are the two-particle states described above, which classically do so, and thus are expected to have probability essentially unity for black hole formation.

This means that a state\foot{A more careful treatment uses narrow wavepackets.} $|E;a\rangle_{in}$ can be rewritten in terms of states that at a time just after formation corresponds to a combined state of black hole and balding radiation; let us choose an orthonormal basis $|E';i\rangle_{rad}$ for the latter, and thus write
\eqn\initstatedecomp{|E;a\rangle_{in} = \sum_{M,I,i}{\cal A}(E,M)_{aIi} |M;I\rangle |E-M;i\rangle_{rad}\ ;}
hereÊ  we neglect the possibility of a small component on states that are {\it not} black holes.
In principle we can project on a definite state of the radiation, yielding a pure black hole state:
\eqn\stateproj{{}_{rad}\langle E-M;i|E;a\rangle_{in} = \sum_I {\cal A}(E,M)_{aIi} |M;I\rangle\ .}

In a generic black hole basis we expect the amplitudes ${\cal A}(E,M)_{aIi}$ to be of order $e^{-S(M)/2}$, corresponding to the fact that from \denstat\ we expect there to be ${\cal O}(e^{S})$ states. The space of states  in \stateproj\ can be combined to form an orthonormal basis for a subspace of black hole states, denoted  $|M;A\rangle$, and labeled by the initial and radiation state labels. However, this basis will not span the space of all black hole states, since \stateproj\ yields too few states.  Indeed, note that there are arguments (extending \refs{\tHooholo}) that only of order 
\eqn\qftstates{\exp\{E^{(D-2)(D-1)\over D(D-3)}\}}
states can be formed from collapse of matter of energy $E$; thus $a$ should have such a range.  If one also accounts for the balding radiation, as above, there are more states that can be accessed through their entanglement with this radiation.   Typical radiated quanta have energies $\sim 1/R$, and  given the radiated energy $E-M$, this yields an entropy $\sim R(E)(E-M)\propto E^{(D-2)/(D-3)}$.  This  exponentiates to give the number of states over which the index $i$ can range.  However, this is still far fewer than the $\exp S(M)$ black hole states, since typically $M>E/2$.  Thus, the number of states that are ``accessible" in the collision at energy $E$ is far less than the number of possible states of the black hole.  We can label a basis for the remaining complementary black hole state space as $|M;{\bar A}\rangle$.  One expects that one approach to accessing these states is to form a black hole of mass $M'> M$ in a higher energy collision,
and then allow it to evaporate down to mass $M$.  In doing so, internal states of the black hole become entangled with the state of the Hawking radiation, like in the preceding discussion of balding radiation.\foot{One can in principle ``purify" such states by projection on definite states of the Hawking radiation, as with the preceding projection of balding radiation.}   For large enough $M'$, this gives 
$e^{S(M)}$ independent accessible states. For many purposes, it is simplest to forget the balding radiation, which as we have explained does not appear to play a particular central role, and in a slight abuse of notation, think of the labels $A$ as corresponding to the initial states from which the black hole formed.

We can likewise label the possible $n$-body out states, representing the final decay products of a given black hole, as $|E, a\rangle_{out}$. In a similar spirit to the preceding discussion, we could choose a basis  of black hole states  labelled by this out-state description.  Again, we expect the matrix elements between the preceding basis and this one to generically have size $\exp\{-S(M)/2\}$.  
 Correspondingly, the amplitude for a given initial black hole state to decay into a given final state of the Hawking radiation  will be of generic size 
\eqn\evapamp{|{}_{out}\langle M,a|M,I\rangle|\sim e^{-S(M)/2}\ .}

The quantum description of black holes as a decaying multi-state system has analogies to other such systems, like  $K0-\overline {K0}$ mesons.  In the assumed unitary dynamics, an initial black hole state $|M;I\rangle$ can both mix with other states with the same energy, and with states that are in the continuum, which consist of a lighter black hole together with radiated quanta.  One might expect, via a Weisskopf-Wigner\refs{\WeWi} approximation, that evolution in the Hilbert space of black hole states with mass $\sim M$ is governed by an effective Hamiltonian:
\eqn\effham{i{d\over dt} |M;I\rangle = H  |M;I\rangle.}
Though conceivably more general dynamics is needed,\foot{In particular, we don't expect $H$ to necessarily be a hamiltonian constructed from a local lagrangian.} this exhibits possible features of black hole evolution.
Due to the decay, the hamiltonian is not hermitian in this subspace, and in general takes the form
\eqn\hmat{H_{IJ} = M_{IJ} -{i\over 2} \Gamma_{IJ},}
where $M_{IJ}$ and $\Gamma_{IJ}$ are hermitian matrices.  In general, these will not commute.

\subsec{Exclusive processes}

\Ifig{\Fig\resonan}{Schematic of a black hole as a resonance in $2\rightarrow2$ scattering.}{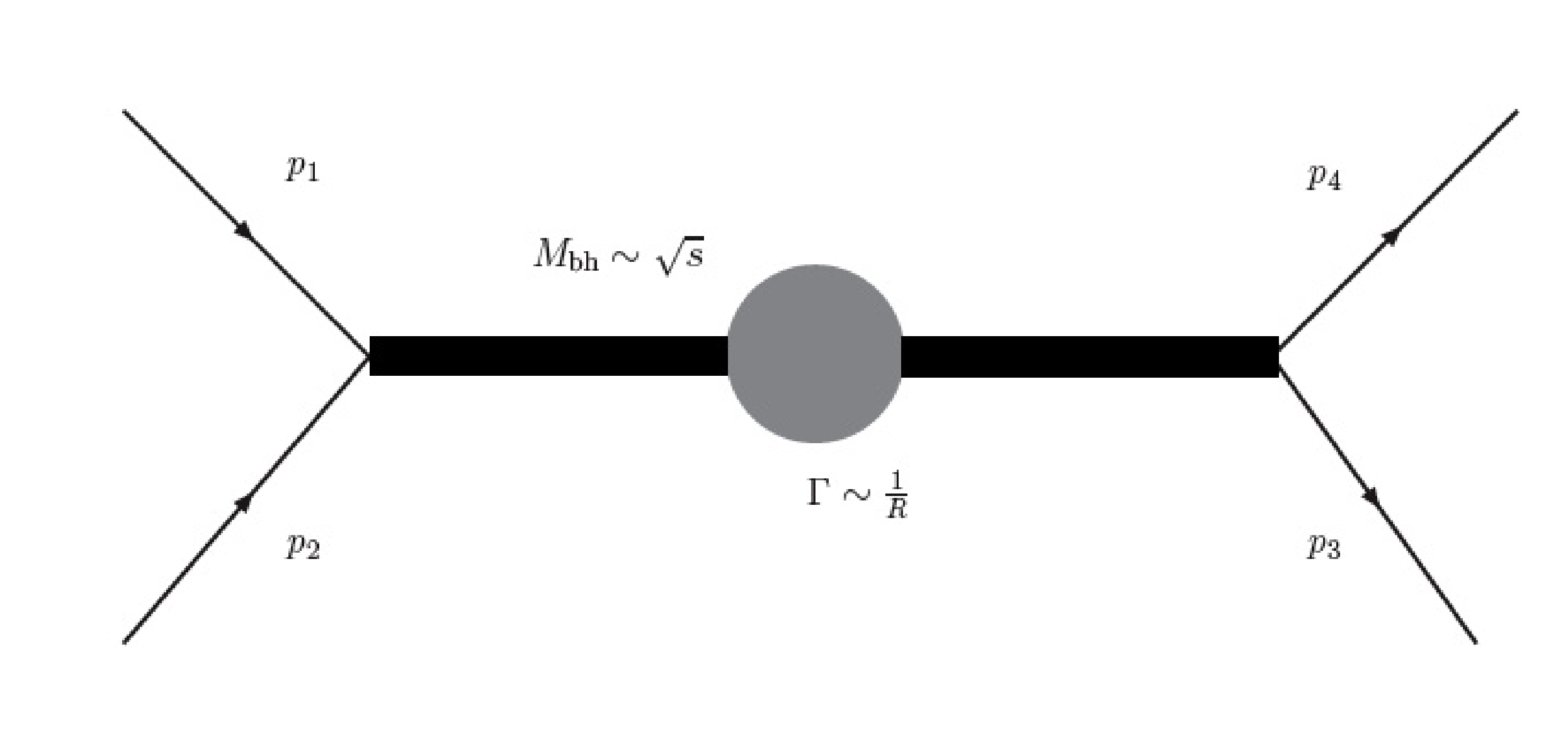}{6}

If one considers in particular an exclusive process with two-particle initial and final states $|p_1,p_2\rangle_{in}$, $|p_3,p_4\rangle_{out}$, such as pictured in \resonan, one thus expects the intermediate black hole states to contribute to the S-matrix as
\eqn\bhevol{
{}_{out}\langle p_3,p_4 |p_1,p_2\rangle_{in}= (2\pi)^D\delta^D(\sum p)
\sum_{IJ} \langle p_3,p_4|J\rangle \left({1\over E-H}\right)_{JI} \langle I | p_1,p_2 \rangle.}
(Note that in the bases adapted to in or out states, described in the preceding subsection, the indices are expected to only range over $\sim S(E)$ values.)
If $M_{IJ}$ and $\Gamma_{IJ}$ do not commute, $H_{IJ}$ cannot be diagonalized by a unitary transformation, but we will assume it can be diagonalized by a more general linear transformation.  The eigenstates $|M;{\check I}\rangle$ are then not orthogonal; 
\eqn\nonortho{\langle M; {\check I} |M;{\check J}\rangle = g_{{\check I}{\check J}}}
for some $g_{{\check I}{\check J}}\neq \delta_{{\check I}{\check J}}$.  In such a basis \bhevol\ becomes\foot{The form of this equation may alternately be simplified through the definition of a dual basis, $\langle \Ic_d| = g^{-1}_{\Ic\Jc} \langle \Jc|$.}
\eqn\diagevol{
{}_{out}\langle p_3,p_4 |p_1,p_2\rangle_{in}= (2\pi)^D\delta^D(\sum p)
\sum_{\Ic\Jc} \langle p_3,p_4|\Ic\rangle {1\over E-H_\Ic} g^{-1}_{\Ic\Jc} \langle \Jc |p_1,p_2\rangle,}
where $H_\Ic = M_\Ic - i\Gamma_\Ic/2$ are eigenvalues.  This will produce a sum of terms of Breit-Wigner form contributing to the amplitude.  However, the sum itself will not, in general, take the Breit-Wigner form.

In the case where the particles being collided are the narrowly-focussed wavepackets that we have described, one plausibly expects the corresponding amplitude to be of size
\eqn\expsuppres{|_{out}\langle a|b\rangle_{in}| \sim e^{-S(E)/2}.}
The reason for this is that for such wavepackets the amplitude to form a black hole is essentially unity, and the amplitude for it to decay back to a two-particle state is of size given by \evapamp.  Note that our discussion suggests a resolution to questions raised\refs{\BaFi} about the relation of intermediate black holes to Breit-Wigner behavior.  One has ${\cal O}(1)$ amplitude to 
 form {\it some} black hole state; in a generic basis for black hole states, this is a superposition with ${\cal O}(e^{-S/2})$ coefficients, although, as indicated in the preceding subsection, one can choose a special basis where black hole states are labeled by the initial states that created them.  Thus, the amplitude to form a generic black hole state from a two-particle state is $\sim e^{-S/2}$, as is the amplitude for a generic black hole state to decay back into a two-particle state.

One might ask whether there could be any larger contributions to the $2\rightarrow2$ amplitude, due to processes that avoid black hole formation.  For example, our gaussian wavepackets will have tails at large impact parameter.  However, these have probability of size $\exp\{-(R/\Delta x)^2\}$ at $b\sim R$.  The width $\Delta x$ is constrained by $\Delta x> 1/E$, but this constraint produces a quantity merely of size $\roughly>\exp\{-S^2\}$.  

While we can't at present rule out other such effects, none have been identified.  Another  test of this statement comes from  scattering of a particle of high energy $E$ off a {\it preexisting} black hole in the relevant range $b\ll R$; here the amplitude $\cal R$ for reflection is also exponentially suppressed\refs{\Sanch}:
\eqn\reflectamp{{\cal R} \sim e^{-4\pi E R}\ .}
It is thus plausible that the amplitude for the {\it classically predicted}\refs{\Penrose,\EaGi,\GiRy} black hole formation process only receives corrections that are exponentially suppressed at least to the level \expsuppres.

\newsec{Partial wave amplitudes}

In this section we restrict attention to $2\rightarrow2$ scattering, in a partial wave basis, and investigate consequences of the preceding picture and related considerations.  For simplicity, we focus on scattering of one species of spinless particles.
The initial two-particle states will be labeled by just their energy and angular momentum $l$, and the scattering amplitude is of form
\eqn\partamp{{\cal S}_l(E) = e^{-2\beta_l(E) + 2i\delta_l(E)}\ .}

\subsec{Strong gravitational regime}

As  outlined above, for impact parameters $b\ll R(E)$, or correspondingly angular momenta $l\ll L(E)$, the amplitude for such a state to form a black hole with total angular momentum $l_{\rm BH} \approx l$ is expected to be of order unity.

\subsubsec{Absorption}

In the $2\rightarrow2$ process that goes through the black hole channel, $l_{\rm BH}=l$.  
From \evapamp, we note that the amplitude for the given resonance $|E,l \rangle$ to decay {\it back} to a two-particle state is $\sim e^{-S(E,l)/2}$. 

As in the preceding section, it is plausible that processes avoiding black hole formation in the regime $l\ll L$ are exponentially suppressed at least to this level.  Arguments for that build on the preceding ones, together with the properties of partial-wave wavepackets.  

For example, consider a wavepacket with definite angular momentum in the relative coordinates between the two particles:
\eqn\pwwave{\psi_{lm}(x) = \int dE {J_{l+\lambda}(Er)\over (Er)^\lambda} e^{-iEt} Y_{lm}(\Omega) f(E),}
where $J_{l+\lambda}$ is a Bessel function, $Y_{lm}(\Omega)$ are $D-2$ dimensional spherical harmonics, and $f(E)$ is a gaussian wavepacket with width $\Delta E$.   Asymptotics of Bessel functions for large order and argument (see eq. 8.41.4 of \refs{\Wat}) then show that for $l\ll Er$, 
\eqn\besselasymp{J_{l+\lambda}(Er) \rightarrow\sqrt{2\over  \pi Er} \cos\left[Er - {\pi(l+\lambda)\over 2}-{\pi\over 4} \right],}
with subleading corrections consisting of terms suppressed by powers of $(l+\lambda)/Er$ times sine or cosine functions of the same form.  Thus combining \pwwave\ and \besselasymp\ gives a wavepacket that is gaussian in $t\pm r$ with width $\Delta r\sim 1/\Delta E$, and subleading terms are similarly gaussian.  

A related argument comes from the relation between the partial wave representation and impact parameter representation\refs{\AdKo}.  Specifically, if $f(b,s)$ is the amplitude in impact parameter representation, then at high-energies one finds the corresponding partial wave amplitude\refs{\CoPe,\AdKounit}
\eqn\hebl{f_l(s) = f(2(l+\lambda)/E,s) + {A\over s} {d^2 f(b,s)\over db^2}\Big\vert_{b=(2(l+\lambda))/E} +\cdots,}
where $A$ is a numerical coefficient,
indicating that in the high-energy limit, localization in angular momentum corresponds to localization in impact parameter, as expected.\foot{The series \hebl\ may be regulated by considering incoming wavepackets instead of plane waves.}

A final argument comes from the behavior of partial waves scattering from a preexisting black hole; \Sanch\ argues that their reflection amplitude in the limit $ER\gg1$ is of size \reflectamp.

Based on these, and on the discussion of section four, we thus conjecture that in the regime $l\ll L(E)$, the $2\rightarrow2$ amplitude is indeed exponentially small in the entropy, and arises mainly due to such a strong gravity channel.  
These statements suggest additional rationale for the ``black hole ansatz" of \refs{\GiSr}, that in this regime
\eqn\absorb{ |{\cal S}_l(E)| = e^{-2\beta_l} \sim \exp\{-S(E,l)/2\} .}
Notice that this behavior has {\it two} characteristic features.  The first is the exponential {\it strength} of the absorption.  The second is the long {\it range} of the absorption, which is characterized by the growth of $L(E)$ with energy.  Even should the preceding arguments regarding the strength of the exponential suppression be evaded, we expect the feature of significant absorption at long range to persist.

\subsubsec{Phase shifts}

We have suggested that the amplitude is essentially unity for a given initial two-particle state with $l\ll L(E)$ to enter the strong gravitational regime.  In $2\rightarrow2$ scattering, one might therefore expect that in each energy range $(E, E+1/R)$ we form one of the  $\Delta\calN(E,l)$ black hole states\foot{As noted, this state is a superposition of states of a generic basis with coefficients of size ${\cal O}(e^{-S/2})$.}  with the corresponding energy and angular momentum.  This would correspond to a density of  ``accessible" states 
\eqn\densacces{\rho_{\rm acc} (E,l) \approx R(E)\ .} 
(This value would be less relevant for $2\rightarrow N$ scattering, where, as we have argued, more states may be accessible and entangle with the balding radiation.)
Notice that this would imply that the {\it total} number of such accessible black hole states of angular momentum $l$ and energy $<E$ is given by
\eqn\totaccess{\calN_{\rm acc}(E) = \int_0^E \rho_{\rm acc}(E,l) dE \approx S(E,l)\ .}

Consider the parametrization \bhevol\ of the contributions of intermediate black hole states.  If the matrix $H_{IJ}$ were diagonal in the ``in"-state basis $|M;A\rangle$, discussed in section four, then we would expect a contribution to the amplitude of Breit-Wigner form:
\eqn\twoamp{e^{2i\delta_l(E)} \approx e^{2i\delta_b}\left(1- {i\Gamma \over E-E_{BH}+i\Gamma/2}\right)\ }
where $\delta_b$ is a ``background" value.
Then, the phase $\delta_l(E)$ would increase by $\pi$ as we pass through each such ``accessible" (or strongly coupled) resonance, and correspondingly, the combined effect of resonances at increasing energies would give
\eqn\phasediag{\delta^{\rm diag}_l(E) = \pi \calN_{\rm acc} (E,l)\approx \pi S(E,l),\ }
as with Levinson's theorem  for single-channel scattering.  Note also that such a result would yield a decay time $d\delta_l/dE\sim R(E)$, compatible with the width $\Gamma\sim 1/R$.

However, we see no reason to expect $H_{IJ}$ to be diagonal, and so consider phase shifts of a more general form, which we parameterize as 
\eqn\phase{\delta_l(E) = \pi k(E,l) S(E,l)}
where $k(E,l)$ varies more weakly with energy than $S(E,l)$.
One might expect $k(E,l)>0$ (corresponding to  time delay) due to the attractive nature of gravity. Indeed, in scattering off a pre-existing black hole the gravitational field introduces a positive phase shift relative to scattering from the angular momentum barrier.  We will investigate additional constraints on $k(E,l)$ in subsequent sections.

To summarize, combining \phasediag, \phase\ suggests that the partial wave amplitudes in the strong gravity regime take the form
\eqn\bhpws{f^{\rm SG}_l(s)\approx {i\over 2} \left\{1- \exp\left( -{1\over 2}S(E,l)[1-4\pi i k(E,l)]\right)\right\}\ .}
Notice that this expression differs from that of \GiSr; that analysis did not take into account the role of inelasticity and accessibility of resonance channels.  Thus \bhpws\ comprises an improvement of the black hole ansatz of \GiSr.

\subsec{Born and eikonal}

One can likewise infer properties of the partial waves in the longer-distance regimes, where the Born or eikonal approximations are expected to be valid.  In particular, ref.~\GiSr\ computed the eikonal phase shift,
\eqn\dbeik {\delta^{\rm eik}_l(E) = {\sqrt{\pi} (D-2) \Gamma[(D-4)/2]\over 8 \Gamma[(D-1)/2] }{L(E)^{D-3}\over l^{D-4}} \sim {E^{D-2} \over l^{D-4}} \ ,}
and checked that the eikonal amplitude unitarizes the Born amplitude, which is the leading term in an expansion in $\delta_l$, as expected.  Thus the transition from Born to eikonal regimes occurs in the small angle regime $l\sim E^{(D-2)/(D-4)}$.  Notice that 
the phase shifts are indeed positive definite, as expected from the attractive nature of gravity.\foot{ This is the case provided $D>4$. The four dimensional case suffers from Coulomb-like singularities, requiring the usual {\it inclusive} amplitudes,  avoided in this paper by working in higher dimensions.}  The correspondence between the eikonal amplitudes and the semiclassical picture\refs{\tHooultra,\ACVone-\ACVfour} suggest the utility of the eikonal description until $l\sim L$.

For decreasing impact parameter/increasing scattering angle, different effects can contribute to absorption.  A generic effect is soft-graviton bremmstrahlung.  This was estimated in \GiSr\ to give a contribution of size 
\eqn\softg{\beta_l^{\rm br}\sim {L(E)^{3D-9}/ l^{3D-10}} \sim {E^{3D-6} \over l^{3D-10}}\ . }
Note that this matches onto the energy dependence of \absorb\ at $l\sim L$, which also fits with a picture where a non-negligible fraction of the collision energy can be emitted in the balding radiation.

As noted in section three, there may be other less-generic effects, e.g. due to excitation of internal degrees of freedom of the colliding bodies.  In string theory, such an effect is the ``diffractive excitation" or ``tidal string excitation" explored in \refs{\ACVone-\ACVfour,\LQGST,\GGM}.  But, as noted, we do not expect such effects to prevent amplitudes from matching onto those of the strong gravitational regime.

\subsec{Combined pictures}

\Ifig{\Fig\betaone}{Absorption coefficients at a fixed angular momentum as a function of the CM energy.}{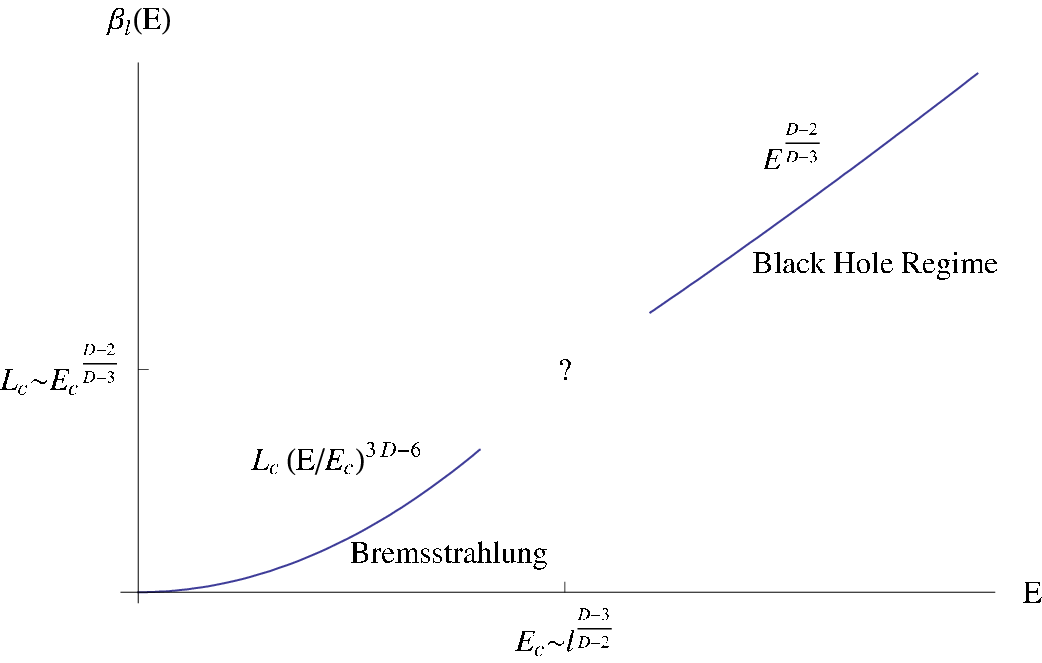}{4}

\Ifig{\Fig\betatwo}{Absorption coefficients at a fixed CM energy as a function of angular momentum, with $L_c \equiv L(E)$.}{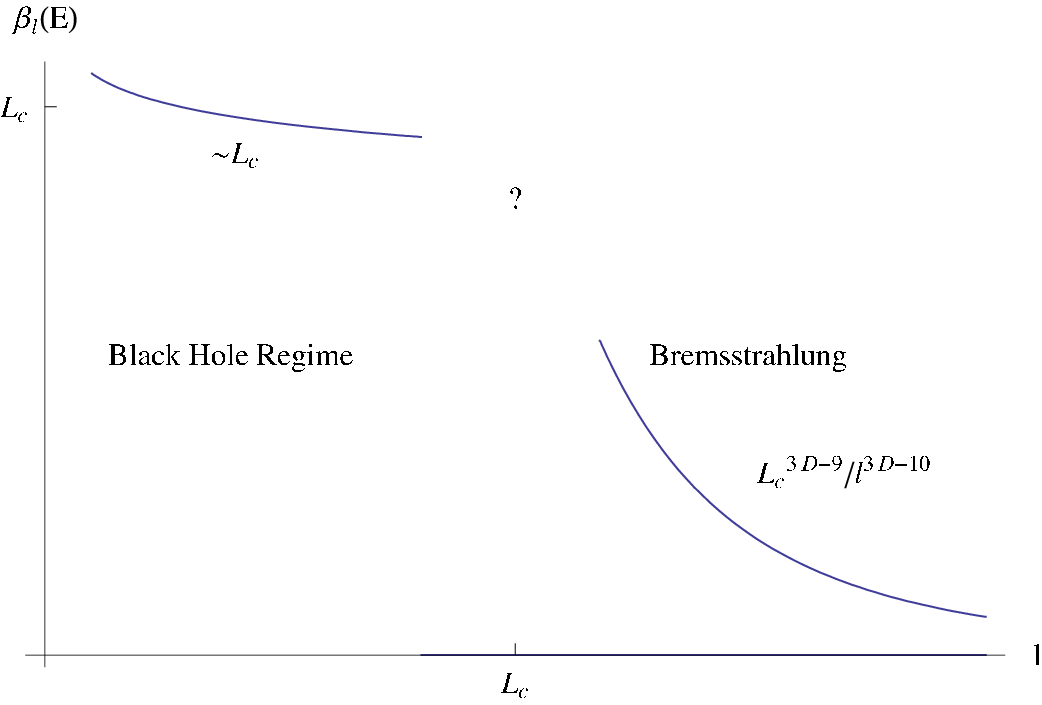}{4}

We can thus suggest combined pictures describing the weak and strong coupling regimes.  The results \softg\ and \absorb\ suggest energy and angular momentum dependences of the absorptive coefficients $\beta_l$ as pictured in \betaone, \betatwo.  

\Ifig{\Fig\figphshone}{Phase shift for fixed angular momentum as a function of the CM energy.}{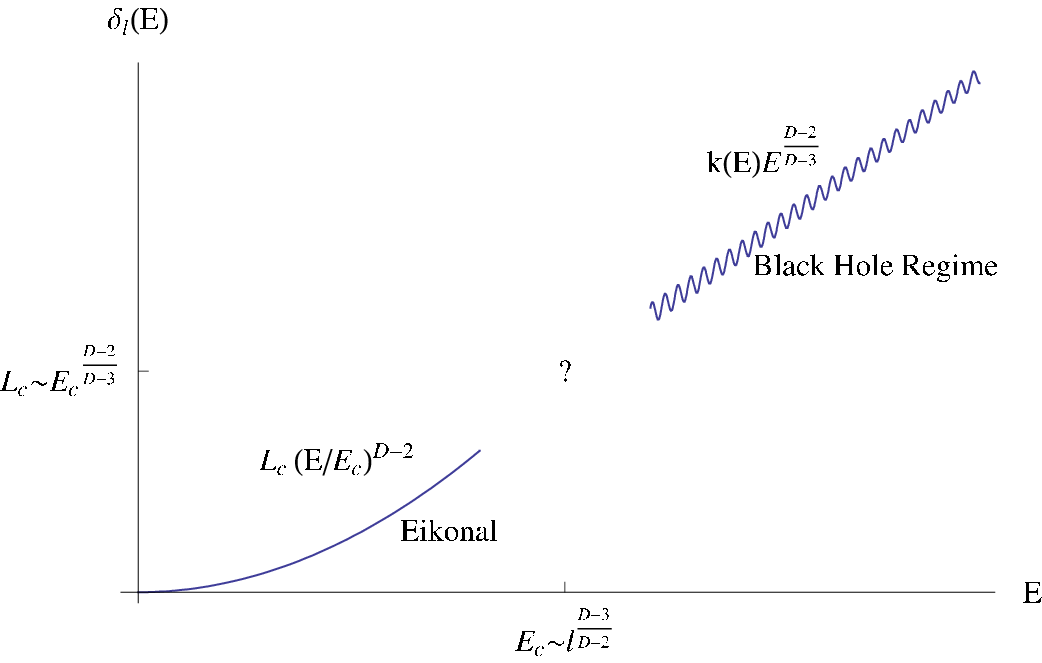}{4}

\Ifig{\Fig\figphshtwo}{Phase shift for a fixed CM energy as a function of angular momentum, with $L_c \equiv L(E)$.}{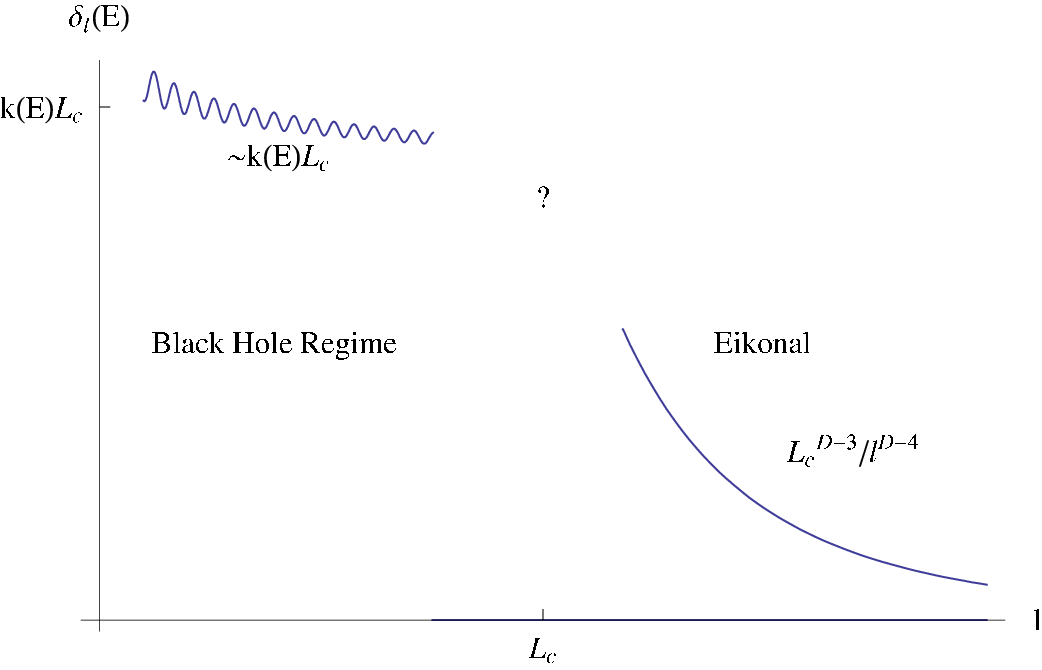}{4}

While the phase shift is well-studied in the eikonal regime, as we have indicated, we have less information in the strong gravity regime, but expect an increase bounded by $\delta_l(E) \sim E^{(D-2)/(D-3)}$ as in \phase.  Sketches of energy and angular momentum dependence are given in \figphshone, \figphshtwo.

\newsec{Momentum space amplitudes}

We now ask what properties of momentum space amplitudes can be inferred from the preceding discussion.  In section two, we noted the collapse of the Lehmann ellipse, and in particular that convergence of the partial wave expansion cannot extend past $t=0$ to positive $t$.  Likewise, continuation of $s$ to complex values with fixed real $t<0$ would correspond to complex $\cos \theta$, outside the convergence region.  These and related limitations restrict our ability to prove results that follow in massive theories.  However, we have argued that the expression for the partial wave coefficients, \pwderiv, is expected to be well-defined and finite.  This means that  properties of the $f_l(s)$ are those of the corresponding integral, and this in turn constrains the behavior of $T(s,t)$.  

Additional information about the momentum space amplitudes comes directly from their eikonal approximation, \Teik.  
At very small angles, this expression reduces to the Born amplitude, \born.  The match between the Born and eikonal regimes occurs near $\chi\sim 1$, corresponding to $t\sim -s^{-2/(D-4)}$ or 
\eqn\borneik{\theta_{B/E} \sim {1\over E^{(D-2)/(D-4)}}.}
The asymptotics of the eikonal amplitude at larger angles follows from performing the integral over angles in \Teik, which yields
\eqn\eikr{iT_{eik}(s,t) = -2is(2\pi)^{(D-2)/2}q_\perp^{-(D-4)/2}\int_0^\infty dx_\perp x_\perp^{(D-2)/2} J_{(D-4)\over 2}(q_\perp x_\perp) (e^{i\chi(x_\perp,s)} -1).}
Then, combining the  Bessel function asymptotics \besselasymp\ with a saddle-point approximation of the integral gives an asymptotic amplitude of the form
\eqn\eikasymp{T_{\rm eik}\sim\exp\left\{ i [s (-t)^{(D-4)/2} ]^{1/(D-3)}\right\}\ .}
This exhibits some interesting features -- such as nonpolynomiality -- that we will return to in the next section.

One may also inquire about implications for $T$ of the strong gravity behavior outlined in the preceding section.  Recall that the physical features of that behavior were 1) significant scattering, and moreover absorption, to an angular momentum that grows with energy as $l\sim L(E)$, 2) strong absorption for large $E$ and $l\ll L(E)$, and 3) potentially rapid growth in the phase, \phase.

For $\delta_l=\beta_l=0$, \pwsoln\ gives $f_l=0$, so the first feature implies nonvanishing $f_l$ to $l\sim L(E)$; significant absorption moreover implies that $f_l\sim i/2$.   These become conditions on the integral
\eqn\pwrew{ \int_0^\pi d\theta\, \sin^{D-3}\theta\, {\cnl(\cos\theta)\over \cnl(1)} T\left[s,t(s,\theta)\right] = {\gamma_D f_l(s) \over s^{(D-4)/2}},}
where $t(s,\theta)$ is given by \tutheta.  However, a direct statement about $T$ in the strong gravity regime $s\sim -t$, is not easily inferred from the significance of the right side of \pwrew, since the integral in particular receives a contribution from the Born regime.  For $\theta<\theta_{B/E}$ and $l<L$, one has $l \theta \ll 1$ and  can use the small-angle approximation
\eqn \gegel {C_l^\lambda (1-\theta^2/2) \simeq C_l^\lambda(1) \left( 1 - {l(l+2\lambda) \theta^2  \over 2(2\lambda+1)}\right)\ . }
The Born contribution to \pwrew\ is thus of size
\eqn\Borncontrib{\int_0^{\theta_{B/E}} d\theta \theta^{D-3} {E^2\over \theta^2} \sim {1\over E^{D-4}}.}
This shows that one expects a contribution to partial wave amplitudes from both the Born and eikonal regions that is significant at angular momenta $l\roughly<L(E)$.

Indeed, a related fact is that the cross-section due to this small-angle scattering is expected to be large as compared to that of the strong gravity region, 
\eqn\SGcross{\sigma_{SG}\approx \pi [R(E)]^{D-2}\sim E^{(D-2)/(D-3)}.}
For $D>6$, where the small angle contribution converges, it can be estimated using the impact parameter where Born and eikonal match, giving\ACVone
\eqn\BEcross{\sigma_{B/E}\sim E^{2(D-2)/(D-4)}.}

Large growth of $\beta_l$ and $\delta_l$ with energy imply that $f_l-i/2$, or $df_l/ds$, are small, and rapidly oscillating.  Eq.~\pwrew\ thus indicates that $T(s,t)$ correspondingly has rapid falloff and oscillations.  Moreover, we see that exponential falloff of $f_l-i/2$ would indicate precise cancellations between the contributions of $T(s,t(s,\theta))$ in the Born, eikonal, and strong gravity regimes; as we have discussed, physical aspects of the scattering such as the analogy with scattering from a fixed black hole suggest such falloff.

A sharper statement arises if one considers continuation of \bhpws\ into the complex $s$ plane.  This form for $f_l(s)$ suggests that generically it would grow exponentially somewhere in the complex $s$ (or $E$) plane.
In particular, for small enough $k$, one finds exponential growth in the $s$  upper half plane (UHP) $0<{\rm Arg} s < \pi$:  for constant $k$, this would occur for
\eqn\kcrit{k< {1\over 4\pi} \tan {\pi\over 2(D-3)},}
and likewise for the example of a decreasing power, $k\propto E^{-p}$.  By \pwrew, this corresponds to exponential, thus not polymomially bounded, growth in $T(s,t)$ for complex $t$.  While with the specific functional form \bhpws, a phase that is too small leads to growth that is not polynomially bounded, it is conceivable that a more complicated analytic structure of the exact amplitude avoids this conclusion.\foot{Though, with added assumptions like hermitian analyticity/dispersion relations, one may possibly generalize methods of \refs{\SuTu,\Kino} to 
show that the exponential falloff in \bhpws\ implies a lower bound on the phase, {\it e.g.} $\delta \roughly> \log s$, given a polynomial bound  in the UHP; also, certain analyticity assumptions together with this falloff might possibly be used to prove violation of polynomial bounds in some region, with methods like in  \refs{\Boas,\Martinb}.  We leave these for future investigation. (Notice that in QFT we do not expect such a strong absorptive behavior, thus polynomial boundedness  is expected to lead to a phase bounded above by $\log s$.)}

\newsec{Analyticity and crossing}

We have investigated aspects of unitarity, particularly via the partial wave expansion; we now turn to analyticity and crossing.

Consider scattering of two massive particles of mass $m$ coupled to gravity.  We might imagine these to be an $e^+e^-$ pair, although to avoid complications of spin we will treat the scalar case.  Another specific context to contemplate, if in a string theory context, is scattering of a $D0-\overline{D0}$ pair.

First, consider behavior for fixed real $t<0$, as a function of $s$.  The two-particle cut in the s-channel begins at $s=4m^2$.  However, one can also have such a pair annihilate to two or more gravitons (in the absence of a net conserved charge), implying multiple cuts beginning at $s=0$.\foot{One might also contemplate the possibility of worse behavior, {\it e.g.} $\sim e^{-1/s^p}$ for some $p$.}  Likewise, there are multiple u-channel cuts beginning at $u=0$.  Given
\eqn\Mansum{s+t+u = 4m^2\ ,}
we find that the u-channel cuts, for fixed $t$, originate at 
\eqn\ucut{s=4m^2-t\ ,}
and are taken to extend along the negative $s$ axis.
Thus, these cuts overlap those from $s=0$ -- there are branch cuts running all along the real $s$ axis, with no gap between them, unlike the massive case.  These features of massless theories weaken some of the constraints present in massive theories.

We likewise expect singular behavior at $t=0$; we have noted the Coulomb pole there, but one might find a more general singularity ({\it e.g.} branch point) when higher-order processes are accounted for.  As we have already described, this prevents the usual continuation along the real axis from $t<0$ to $t>0$, that is a useful tool in massive theories.

\subsec{Crossing symmetry}

For real $s_0>4m^2$, the physical amplitude with $s=s_0$, $t<0$ is assumed to arise from the analytic function $T(s,t)$ with $s=s_0+i\epsilon$ in the limit $\epsilon\rightarrow0^+$. By the maximal analyticity hypothesis, $T$ only has singularities dictated by unitarity, so can be continued throughout the $s$ UHP; likewise for fixed $s$, one can continue in $t$, avoiding singularities.

In a massive theory, at small $t<0$, one can continue in $s$ across the real axis, through the gap between the cuts.  This allows one to define the amplitude for $s=s_1-i\epsilon$, for large negative real $s_1$, which by \Mansum\ corresponds to u-channel kinematics.  Crossing symmetry is the assumption that a single function $T(s,t,u)$, with variables satisfying \Mansum, defines amplitudes in all channels through such continuation.

Clearly this specific continuation fails in the massless case, given the lack of a gap between the cuts.  However, it appears possible to still obtain crossing, through use of a different path.

\subsubsec{The BEG path}

Such a path was given by Bros, Epstein, and Glaser in \refs{\BEGtwo}, as follows.  First, begin at large $s_0>0$, and 
hold $u=u_0<0$ fixed. One can continue through the upper $s$-plane to $e^{i\pi } s_0$.  Here, $t$ will approach the positive real axis with a $-i\epsilon$; we can denote this as the $t^-$ channel.   Next, beginning at this point, keep $s<0$ fixed and continue $t\rightarrow e^{-i\pi} t$.  This is analogous to the preceding continuation, and takes $t^-$ to $u^+$ -- here the positive real $u$ axis is approached from above.  The combined path thus continues from the physical s-channel $s^+$ to the physical u-channel $u^+$, permitting crossing.\foot{Note that one must also include a small path segment from $(s,t,u)=(-s_0+i\epsilon, 4m^2 -u_0 + s_0 -i\epsilon, u_0)$ to $(-s_0, 4m^2 -u_0+s_0-i\epsilon, u_0 +i\epsilon)$.  We assume this is permitted by sufficient holomorphy in this neighborhood, as in \refs{\BEGone}, though more systematic investigation is conceivably warranted.}

\subsec{Crossing and polynomial boundedness}

\Ifig{\Fig\figPL}{The complex $s$ plane, indicating some of the relationships entering into the Phragmen-Lindel\" of argument for a polynomial bound.}{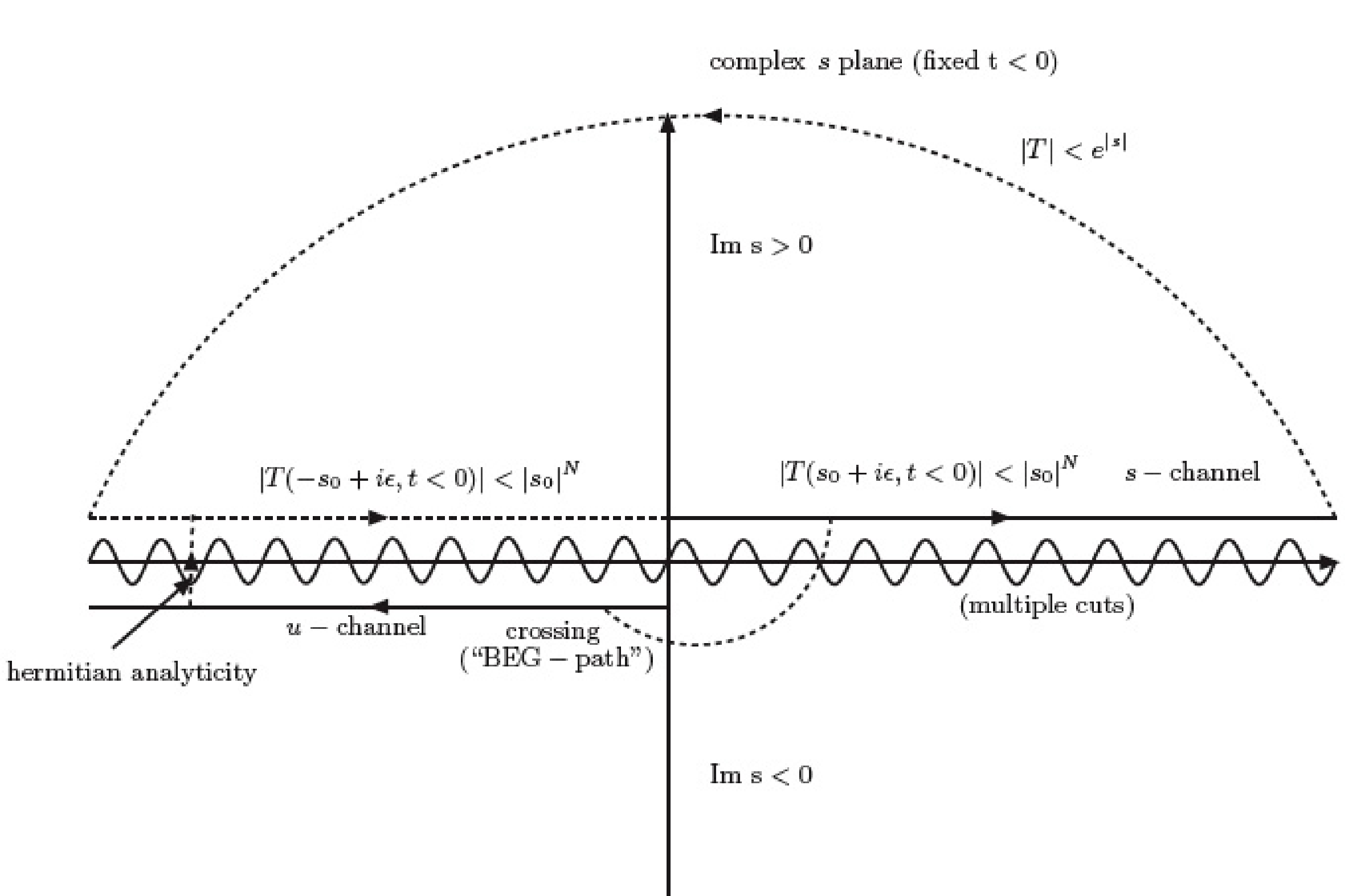}{6}

Analyticity and crossing constrain possible non-polynomial behavior, as we will now discuss; the reader may wish to refer to figure \figPL.  This observation follows from 
the  Phragmen-Lindel\" of Theorem:
{\it If an analytic function is bounded along two straight lines sustaining an angle $\pi\over \alpha$, e.g. $|T(|s|)|<M$ on the lines, and if $T(s)$ grows at most like $e^{|s|^\beta}$ with $\beta < \alpha$ in any other direction, then in fact $T(s)$ is bounded by $M$ in the whole sector sustained by the two lines.}

Choose, for example, $\alpha=1$.  Let us assume that the amplitude is quite weakly bounded, $|T(s,t<0)| < e^{|s|}$.  Note that this bound is easily satisfied both by the eikonal behavior \eikasymp, and by behavior that could arise from growth of the strong gravity region, either from the large absorption coefficients $|\beta_l(s)| \sim |s|^{(D-2)/(2(D-3))}\ll |s|$, or the large range $R(E)\sim E^{1/(D-3)}$  which suggests behavior\GiSr\ (see the next section), $T(s,t<0)\sim e^{R(E)\sqrt t}$. Therefore, by the theorem, if we had a non-polynomial growth in the UHP, that would also require a non-polynomial growth in a straight line $i\epsilon$ above the real axis from $-\infty$ to $+\infty$. 
 
The region $[0,+\infty)$ corresponds to the $s$-channel amplitude.  However,  properties of the Gegenbauer polynomials combined with the optical theorem (see appendix) show that ${\rm Im} T(s,t<0) < {\rm Im} T(s,0) \sim s \sigma_T(s) < s^N$.  (The polynomial bound at $t=0$ is directly connected to existence of a forward dispersion relation\GiSr, following from causality, to be discussed in the next section.)  Moreover, we have the high-energy expression
\eqn\PLreal{\int_0^\pi d\theta\sin^{D-3}\theta |{\rm Re} T|^2 \propto \int d\Omega_{D-2} |{\rm Re} T|^2 < \int d\Omega_{D-2} |T|^2 \propto s^{3-D/2} \sigma_{2\rightarrow 2} < s^{3-D/2} \sigma_T,}
where proportionality is modulo numerical coefficients, and therefore the real part of the amplitude also must be polynomially bounded, provided it is sufficiently smooth.  (Recall that in the strong gravitational regime the real part of the amplitude is indeed subdominant due to strong absorption).
 
 In massive theories, the  $(-\infty,0]$ region is related to the $u$ channel amplitude by complex conjugation.\foot{A rough argument for this follows from the relation between the continuations $s\rightarrow-s$ and $E\rightarrow-E$;  the latter corresponds to taking the complex conjugate of the amplitude.}  This follows from the property of {\it hermitian analyticity} or extended unitarity, which is the requirement $T(s^*,t^*)=T(s,t)^*$. Notice that this implies $f_l(s^*)=f_l(s)^*$ for the partial wave coefficients. 
If we work at negative values of transfer momentum, e.g. $t<0$, hermitian analyticity also connects the discontinuity across the cuts due to threshold singularities to the imaginary part of the amplitude by
\eqn\discT{ {\rm Disc} T (s,t) = 2i {\rm Im} T(s+i\epsilon,t)\ .}
With a mass gap, hermitian analyticity follows from reality of the amplitude below threshold, along with the Schwarz reflection property.  
 In massless theories the status of hermitian analyicity remains unclear, although it seems to hold at any order in perturbation theory. 
If hermitian analyticity holds in gravity, it thus also forbids non-polynomial growth along $(-\infty,0]$, and so by the above theorem, in the UHP of  $s$.

A conservative conjecture is that gravity respects both crossing symmetry and hermitian analyticity, and that amplitudes thus satisfy such a polynomial bound.  We can check this in the asymptotics of the eikonal, \eikasymp, which does so for $D>4$, as does the preceding strong gravity expression.

Nonpolynomiality of amplitudes is however generally expected to give unbounded behavior in other regions of $s,t$, and $u$.  Indeed, one can directly see indications for such behavior given the partial wave coefficients \bhpws.  For example, if $k(E,l)\sim E^{-p}$ for some $p>0$, then the strong-gravity $f_l$'s given by \bhpws\ will have polynomially-unbounded behavior somewhere in the UHP Im$(s)>0$.  Then, \pwrew\ implies that $T[s, t(s,\theta)]$ must likewise be unbounded.  Notice, though, that this is for fixed $\theta$ rather than $t$; thus unboundedness at large $|s|e^{i\phi}$ corresponds to $t\sim - |s|e^{i\phi}$.  As discussed, even $k(E,l)= {\cal O}(1)$ does not necessarily eliminate this behavior, though positive $k$ -- corresponding to time delay -- decreases the region of non-bounded behavior in the UHP.  Likewise, $k<0$, corresponding to a time advance, increases the domain of this behavior.  One also observes unbounded behavior from the eikonal phases, \dbeik.

It is interesting that a polynomial bound in the physical region Im($s)>0$, $t<0$ (and correspondingly in other channels) follows from the very general assumptions that we have described, together with the assumption of causality in the form of the forward polynomial bound.  We next turn to investigation of connections between polynomiality and locality.

\newsec{Locality vs. nonpolynomiality}

The status of locality in gravity is a very important question, given that it is one of the cornerstones of a local quantum field theory description of nature.  Locality is 
also one of the assumptions leading to the information paradox, and conversely, certain violations of locality inherent to nonperturbative gravity have been proposed as the mechanism for information to escape an evaporating black hole\refs{\GiLi,\LQGST,\GiddingsSJ,\QBHB}.\foot{For earlier proposals of a role for nonlocal effects, see \refs{\BHMR,\tHooholo,\Sussholo}.}

If one is restricted to an S-matrix description of dynamics, one can ask how specifically locality is encoded in that description.  In particular, nonpolynomial behavior in the momenta, such as we have described, is suggestive of non-local behavior;\foot{Although, formulations of local field theory with mild nonpolynomial behavior have been proposed\Jaff.} a first heuristic for this is the observation that  nonpolynomial interactions take the form $e^{\partial^n}$ in position space, which is clearly not local.

For massive theories, sharper statements can be made.  In particular, commutativity of observables outside the lightcone can be used to show that the forward amplitude is polynomial bounded\refs{\GGT}, $|T(s,0)|<s^N$.  With a mass, such statements can be extended\refs{\Martext}  both to $t<0$ and to complex values of $t$, including $t>0$.

Diffeomorphism invariance forbids local observables in gravity. 
It has been proposed that local observables are approximately recovered from certain relational {\it protolocal observables};  initial exploration of them in effective field theory is described in \refs{\GMH,\GaGi,\GiMa}.  However, as yet no sharp criterion for locality can be formulated in terms of these observables, and indeed it has been argued\refs{\GiLi,\GMH} that there are fundamental obstacles to such precise locality.\foot{For further discussion, see \GiddingsPJ.}

Nonetheless, bounds on amplitudes can also be understood from a physical perspective, in connection with causality.  This becomes particularly clear with forward scattering.  

Consider first $0+1$ dimensional scattering, with initial and final amplitudes related by an S-matrix,
\eqn\Stime{\psi_f(t) = \int_{-\infty}^\infty dt' S(t-t' ) \psi_i(t').}
Causality states that if the source $\psi_i$ vanishes for $t'<0$, the response $\psi_f$ does as well.  In the complex energy plane, this arises as a result of $S(E)$ having the appropriate analytic structure, and in particular the needed contour deformation arguments require that $S(E)$ be polynomially bounded in the UHP for $E$.  For example, $S(E) = e^{-iE\tau}$ would produce an acausal time {\it advance} by $\tau$. 

The arguments for higher-dimensional forward scattering can be formulated in analogous fashion; a wavepacket that scatters at zero angle should not reach infinity more rapidly than one that does not scatter, implying a polynomial bound, and corresponding dispersion relations.\foot{The relations between causality, analyticity and a well defined UV completion are interesting and subtle. Indeed, other strong restrictions on which IR behavior can be consistently completed into a causal UV theory, given existence of forward dispersion relations, are described in  \refs{\adams,\DGPP}.}
Whereas in the massive case such a bound also implies bounds for $t\neq0$, the collapse of the Lehmann ellipse that we have noted in the massless case obstructs such arguments. 

\Ifig{\Fig\repulsR}{Illustration of scattering by a repulsive interaction of range $R$; the scattered wave at angle $\theta$ has a path that is shorter by $2R\sin{\theta\over 2}$ relative to a wave traveling unscattered through the origin, thus has a relative time advance.}{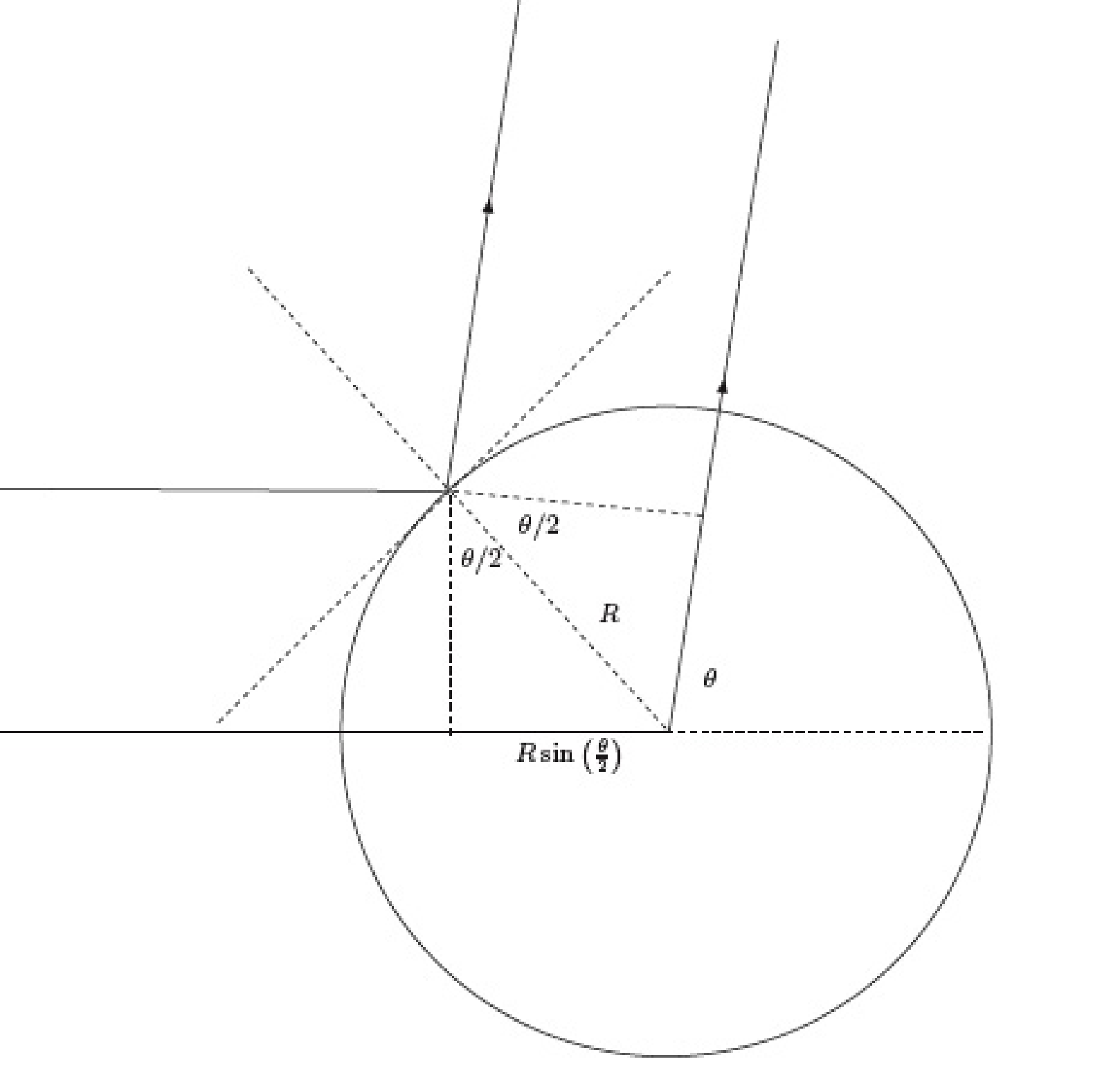}{5}

\Ifig{\Fig\ASpic}{Illustration of scattering of a particle by the gravitational field of an ultrarelativistic source; the scattering angle is negative, corresponding to attraction, and this results in a path for the scattered wave that is longer by $R\sin\theta\sim 2R\sqrt{tu}/s$ as compared to a wave that passes through the scattering center.}{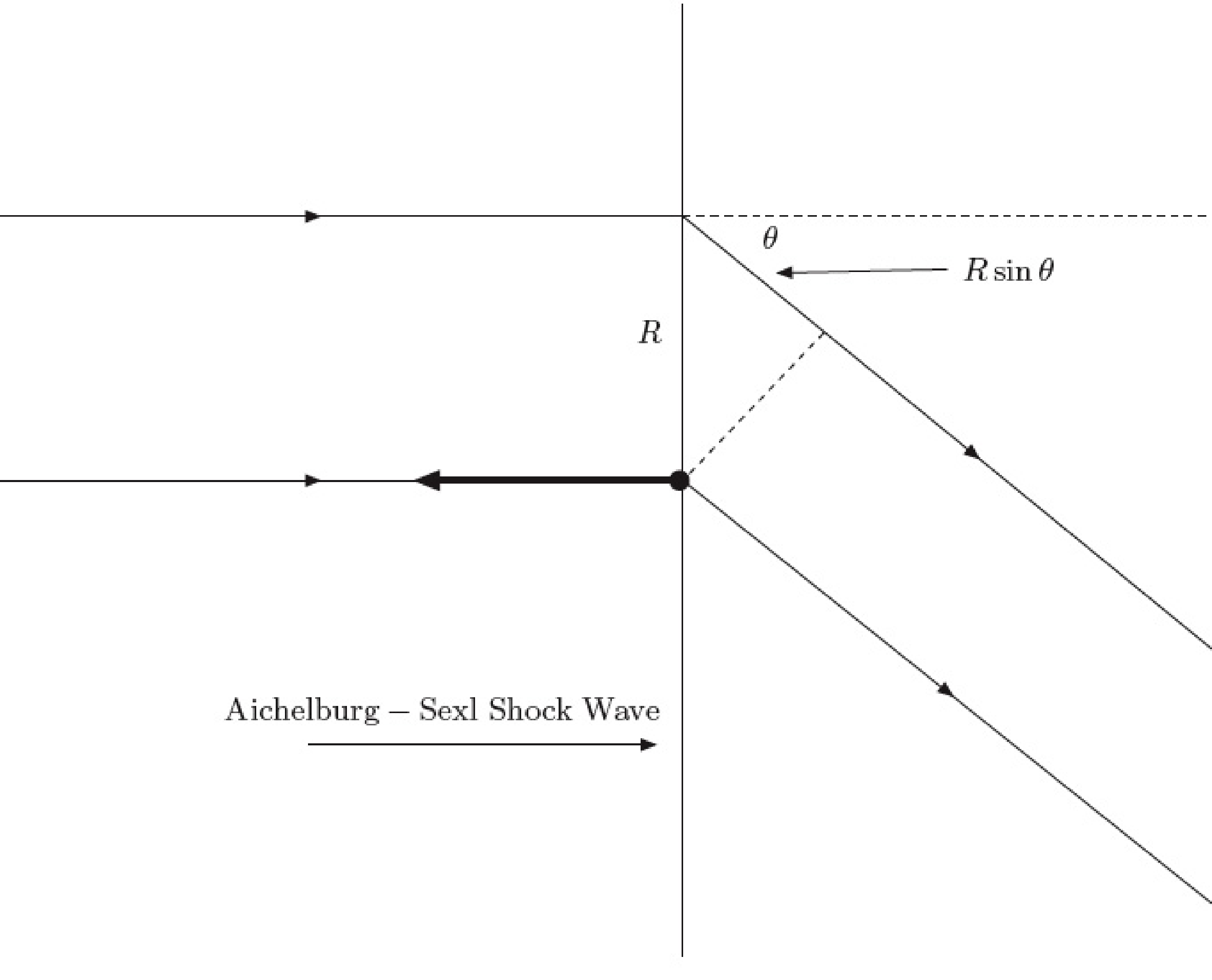}{5}

Consider, however, a physical picture of non-forward scattering, as described in {\it e.g.} \refs{\Newt}; see \repulsR.  If the scattering has a range $R$, a wavepacket can shorten its path by an amount up to $R|q|/E$ with respect to a path going through the origin, with a corresponding time advance.  Thus, we would expect asymptotic behavior
\eqn\srepuls{S\sim e^{-i\sqrt{-t}R}}
which is {\it not} bounded.  Note, however, that such a picture is appropriate to a {\it repulsive} potential.  If one instead considers scattering in gravity, {\it e.g.} in the background of a high-energy particle, whose gravitational field is approximately Aichelburg-Sexl (see \ASpic), the scattering angle is negative, and  the particle receives a time delay, corresponding to positive phase shift, appropriate to an attractive force.  If of finite range $R$, this corresponds to behavior
\eqn\sattr{S\sim e^{i\sqrt{-t}R}.}

In this way, long range behavior of this kind, which in the absence of a better definition we will also call nonlocal, does not obviously conflict with causality.  The danger of a conflict appears even less in an attractive case which produces only time delays; correspondingly one has a polynomial bound for $R\propto E^p$ in this case when $E$ undergoes a small enough positive phase rotation.  Thus, plausibly, nonlocality with time delays is consistent with the existence of a polynomial bound in the physical region, $t<0$, Im$(s)>0$.  The preceding section also argued that crossing, hermitian analyticity, and causality imply such a bound.   While the large phase shifts and strong absorption up to large impact parameters that we have inferred on physical grounds might have violated such a polynomial bound in the physical region, we have found no evidence for such behavior.
It remains possible that an exponential growth may emerge at fixed (real) scattering angle, other than $\theta=0$. This however does not seem to contradict any fundamental property we know, but is another possible signal of nonlocal behavior.\foot{As noted, one might also consider the possibility, which we haven't been able to rule out, that amplitudes, while nonpolynomial, may have  sufficiently complicated analytic structure  to stay polynomially bounded in other regions as well.}

In saying this, we should address arguments of  \GiSr\ suggesting behavior combining \srepuls\ with \sattr, where $R=R(E)$, which would be naturally interpreted in terms of a time advance.  However, this arose from a sharp cutoff in the partial wave sum and does not account for the phase shifts.  If one avoids $\theta=0$, where causality requires cancellations of non-polynomial behavior\GiSr, we can write 
\eqn\Tnza{T(s,t)\propto \sum_{l=0}^\infty (l+\lambda) C_l^\lambda(\cos\theta) e^{2i\delta_l(s) - 2 \beta_l(s)}}
(the sum of $i/2$ generates a $\delta(\cos\theta-1)$).  Plausibly, the exact phase shifts and absorptive coefficients yield only time-delayed behavior, and a bound in the $s$ UHP.

In the preceding section, we argued that the effective range of the interaction grows with $E$;  $R\sim E^p$, with $p=1/(D-3)$ for the strong gravity region, and the rough estimate $p=2/(D-4)$, from \BEcross, for the eikonal amplitudes.  It is interesting to compare this behavior to what is commonly regarded as another indicator of unitary local behavior, the Froissart bound, which states
\eqn\rfroiss{R\leq R_f= a \log E}
for constant a.  In a massive theory, there is a direct connection between this bound and polynomial boundedness.  Heuristically, this is seen via 
\eqn\froissexp{e^{\pm R_f \sqrt{-t}}\sim E^{\pm a \sqrt{-t}},}
which is polynomial behavior.  More sharply, the polynomial bound is used directly in the proof of the Froissart bound\refs{\Martbd,\CFV}.  However, this proof proceeds via the partial wave expansion in the region $t>0$, which we have argued is divergent for gravity. 

 It is tempting to conjecture that there is such a direct connection between power-law growth of the cross section in gravity and nonpolynomiality, perhaps through appropriate regulation of the partial wave expansion.  Indeed, as discussed in \GiSr\ and above, the appearance of strong absorption to $L\sim E^{(D-2)/(D-3)}\gg E\ln E$ implies nonpolynomial behavior of a truncated partial wave sum.\foot{Note that such strong absorption directly corresponds to a cross-section with growth \SGcross.  This follows from taking $\beta_l\gg1$ for $l\ll L$ in \pwexp\ evaluated at $\theta=0$; this, together with the large-$l$ asymptotics $C_l^\lambda(1)\sim l^{2\lambda-1}/\Gamma(2\lambda)$ gives $T(\theta=0)\sim i s^{(4-D)/2} L^{D-2}$, and thus, by the optical theorem, \SGcross. Of course, as we have noted, an even larger contribution to $\sigma_T$ comes from the eikonal region.}  However, as we have argued, we expect the full sum to be polynomial bounded in the $s$ UHP, even if it is not polynomial. One issue arising from massless modes is that we cannot neglect the tail of the partial wave expansion, as one does for example in theories with a mass gap, where $f_l$ decays exponentially for $l \gg E \log E$.  In our gravitational context, these large impact parameter contributions are central in producing the IR singularities at $t=0$.  Indeed, masslessness also plays an important role in the form of the amplitudes in the eikonal regime  (where $l\gg L$), which appears to dominate the cross-section at large energies.
Since the partial wave expansion does not converge at $t>0$, the Froissart bound can be violated without collateral damage. We may associate this with a sort of IR/UV mixing, in the sense that the singularities in the IR (correspondingly the long-range character of gravity) permit a much faster growth in the cross section deep in the UV without conflicting with any other fundamental property. Notice that the eikonal amplitudes already provide us with such an example, without explicit reference to the strong gravity region.

One thus finds that masslessness, and in particular singular behavior at $t=0$, nonpolynomiality, and polynomial growth of cross sections are intricately entwined.  One might question whether all novel features follow from masslessness alone.  However, given that one does not find power law growth $R\sim E^p $ in gauge theory, gravity appears distinctive, due in part to the power-law growth of its coupling with energy.  One might conjecture that a massless theory like QED is on the borderline of locality, but gravity is in a real sense not local, as for example evidenced by its growth of range.   Such a conjecture is certainly permitted without a sharper characterization of locality.

It is interesting to consider one known approach to regulating IR behavior in gravity, namely working in an AdS background.  With AdS curvature ${\cal R}\sim \mu^2$, the graviton effectively has a mass $\sim \mu$.  Correspondingly, growth of black hole radius with energy stops being power law once $R\sim 1/\mu$, and one in particular finds evidence for Froissart-like behavior, $R\propto \log E$, for scattering above this energy\refs{\SGfroiss}.  One might likewise expect restoration of polynomial scattering amplitudes.
However, the matter of extracting the S-matrix in AdS remains an open question\refs{\GaGiAds}, despite some recent progress\refs{\GGP,\Heems}.

It is very interesting that no fundamental inconsistency has yet arisen between the conditions of  unitarity, analyticity, crossing symmetry, causality, and nonlocality in the sense described, despite the existence of nontrivial constraints arising from their combination; it is also moreover interesting that gravitational amplitudes could well run the gauntlet among these conditions. This would also been in harmony with arguments that local field theory breaks down in contexts described by the locality bound\refs{\GiddingsSJ,\GiLi,\GiddingsBE}, and with more general statements that the nonperturbative physics that unitarizes gravity (and specifically leads to unitary black hole decay) is not intrinsically local\refs{\GiddingsSJ}, yet retains certain analytic features and aspects of causality  -- particularly those necessary for consistency!  In any case, further exploration of properties of consistent quantum-mechanical amplitudes for gravity is certainly of great interest.

\bigskip\bigskip\centerline{{\bf Acknowledgments}}\nobreak
We wish to thank N. Arkani-Hamed, H. Epstein, M. Green, D. Gross, A. Martin, J. Polchinski, M. Srednicki, R. Stora, D. Trancanelli, G. Veneziano, and E. Witten for valuable discussions.  We greatly appreciate the stimulating hospitality of the CERN theory group over the course of part of this work.
This work  was supported in part by the Department of Energy under Contract DE-FG02-91ER40618,  and by grant  RFPI-06-18 from the Foundational Questions Institute (fqxi.org).  

\appendix{A}{Optical theorem in $D$ dimensions}

From the unitarity of the $S$-matrix we have

\eqn\unit{ T_{\alpha\beta} - T^*_{\beta\alpha} = i \sum_N \int (2\pi)^D d\Phi_N T_{\alpha N} T^*_{\beta N}
}
where we take $\alpha,\beta$ to be the initial and final two-body states with $p_\alpha \equiv p_1+p_2, p_\beta=p_3+p_4$, and the sum runs over all possible $N$-particle states allowed by the symmetries and conservation of energy and momentum. Here we use the  Lorentz invariant normalization of states,
\eqn\normalz { \langle k|k'\rangle = (2\pi)^{D-1} 2\omega_{\bf k}\delta^{D-1}({\bf k}-{\bf k}')\ }
with $\omega_{\bf k}^2 = {\bf k}^2+m^2$, and  introduce the Lorentz invariant measure
\eqn \lime { \widetilde{dk} \equiv {d^{D-1}k \over (2\pi)^{D-1} 2\omega_{\bf k}}\ .}
 If the intermediate $N$-particle state consists of momenta $q_i$, the N-body phase space is defined by
\eqn\phasfac{d\Phi_N = \delta^D\left(p_\alpha -\sum_i^N q_i\right) \prod_{i=1}^N \widetilde{dq}_i
}
Using these conventions we have for the dimensions of the $2 \to 2$ scattering amplitude, $[T(s,t)]= M^{4-D}$.

If we now restrict \unit~to forward scattering, e.g. $\alpha=\beta$, we can replace the LHS by $2i~{\rm Im}T(s,0)$, and on the RHS we recognize the sum of the square of the amplitudes which enters in the definition of the total cross section. Recall that this is defined as
\eqn\cross{ \sigma_T \equiv \sigma(\alpha\to {\rm all}) =  
\left[{1\over 4\sqrt{(p_1\cdot p_2)^2 - m_1^2 m_2^2} }\right](2\pi)^D \sum_N \int d\Phi_N |T_{\alpha N}|^2\ .} 
Notice that the prefactor in square brackets goes to $1/(8E_1E_2)$ when $s\gg m_1^2, m_2^2$.  
We are now ready to state the optical theorem, which is nothing but a direct consequence of unitarity:
\eqn\opt{ {\rm Im}T(s,0) = 2 \sqrt{(p_1\cdot p_2)^2 - m_1^2 m_2^2}~ \sigma_T(s) \to s \sigma_T(s)\ .}
We can also relate the coefficients in the partial wave projections \pwexp, where the optical theorem takes the form (in the $s \gg m^2$ limit) \Sold
\eqn\pwopt {{\rm Im} f_l(s)= 8 (2\pi)^{2D-2} \left({s \over 4}\right)^{2-D/2} \sum_N \delta^D(p_N-p_\alpha) |f_l(s,\{N\})|^2 \ ,}
from which \pwunit~follows. In this expression the $f_l(s,\{N\})$ are the partial wave projections of the generic intermediate states, considered modulo an overall rotation.
The sum runs over all possible such subclasses  of states\Sold. Performing the sum over $l$ on both sides reproduces the optical theorem. 

As we emphasized in this paper, due to the masslessness of gravity we expect singularities at $t=0$. We noticed before that the IR singularities can be removed by working in $D >4$. From the definition of the cross section we promptly discover that we actually need even higher $D$ for it to be well defined. This follows from the elastic cross section;    \born\ gives probability
\eqn\ircross{ |T|^2 \sim {1 \over \theta^4} \ .}
 This Rutherford-like singularity is tamed for $D>6$ by the integration over solid angle, with measure  $\sin^{D-3}\theta$,  giving a finite cross section.
Once the cross sections are finite the optical theorem \opt\ shows that ${\rm Im} T(s,0)$ is also finite.  One may be tempted to push the partial wave expansion to $t>0$, but this attempt fails once we realize that $t=0$ is indeed also a threshold for graviton production, and the partial wave expansion will not converge past that point. The finiteness of ${\rm Im}T(s,0)$ is due to the fact that in higher dimensions the threshold behavior scales as a  power of momentum, e.g. $\sim {(-t)}^{\alpha}$, rather than logarithmically as we are used to encountering in four dimensional field theories. This is intimately linked to the softness of the IR divergences in $D >4$ due to the promotion of the measure in the loop integrals from ${ d^4q \over (2\pi)^4}$ to ${ d^Dq \over (2\pi)^D}$. It is then easy to see that the expansion of the derivatives of $T(s,t)$ at $t=0$ will not converge and we cannot analytically continue the partial wave decomposition to positive values of $t$.

A final comment is in order. The reader may be puzzled by the fact that the Born approximation in \born\ seems to have a divergent imaginary part as $t\to 0$ from the $i\epsilon$ prescription. A careful analysis shows  that is indeed not the case, and such singularity only arises in the plane-wave limit and disappears as soon as we take into account wave packets. The real part of the amplitude is large, but finite, and give rise to a finite contribution in the cross section as in \ircross .

\listrefs
\end